\documentclass{aa}  

\usepackage{hyperref}
\usepackage{graphicx, subcaption, color, float}
\usepackage{xcolor}
\definecolor{darkblue}{rgb}{0,0,0.5} 
\usepackage{transparent}
\usepackage[adobe-utopia]{mathdesign}
\usepackage{amsmath, amssymb, bm, gensymb}
\usepackage{mathtools}

\allowdisplaybreaks[3]

\newcommand{\bs}[1]{\ensuremath{\boldsymbol{#1}}}
\renewcommand{\vec}[1]{\ensuremath{\mathbf{#1}}}

\newcommand{\x}{\ensuremath{\vec{x}}}

\newcommand{\X}{\ensuremath{\vec{X}}}

\newcommand{\y}{\ensuremath{\vec{y}}}

\renewcommand{\iota}{\textit{i}}

\newcommand{\prob}[1]{\ensuremath{P\left(#1\right)}}

\newcommand{\review}[1]{{{#1}}}

\begin{document}

\title{A probabilistic approach to direction-dependent ionospheric calibration}

%Direction-dependent, tomographic and probabilistic calibration of ionospheric effects
%... for low-frequency radio astronomy??

%Probabilistic tomography of the ionosphere for direction-dependent calibration
%... in low-frequency radio astronomy???

%A probabilistic, tomographic approach to direction-dependent calibration of ionospheric effects

%Tomographic
%Direction-dependent
%Ionospheric
%Probabilistic

%Mapping the probabilistic measure of ionospheric free electron density to differential total electron content
%Propagating the stochasticity of ionospheric free electron density to differential total electron content
%Direction-dependent ionospheric calibration with Gaussian processes
%A statistical foundation
%A mathematical foundation

\author{J.~G.~Albert\inst{1} \and M.~S.~S.~L.~Oei\inst{1}\and R.~J.~van~Weeren\inst{1}  \and H.~T.~Intema\inst{1, 2} \and H.~J.~A.~R\"ottgering\inst{1}}

\institute{1. Leiden Observatory, Leiden University, P.O. Box 9513, 2300 RA Leiden, the Netherlands\\
2. International Centre for Radio Astronomy Research -- Curtin University, GPO Box U1987, Perth WA 6845, Australia\\
E-mail: \href{mailto:albert@strw.leidenuniv.nl}{albert@strw.leidenuniv.nl} \& \href{mailto:oei@strw.leidenuniv.nl}{oei@strw.leidenuniv.nl}}
%E-mail: \href{mailto:albert@strw.leidenuniv.nl}{\textcolor{blue}{albert@strw.leidenuniv.nl}} \& \href{mailto:oei@strw.leidenuniv.nl}{\textcolor{blue}{oei@strw.leidenuniv.nl}}}
%E-mail: \href{mailto:albert@strw.leidenuniv.nl}{\texttt{albert@strw.leidenuniv.nl}} \& \href{mailto:oei@strw.leidenuniv.nl}{\texttt{oei@strw.leidenuniv.nl}}}

\date{\today}
\date{Received 12 April 2019 / Accepted 10 October 2019}

\abstract{Calibrating for direction-dependent ionospheric distortions in visibility data is one of the main technical challenges that must be overcome to advance low-frequency radio astronomy.
In this paper, we propose a novel probabilistic, tomographic approach that utilises Gaussian processes to calibrate direction-dependent ionospheric phase distortions in low-frequency interferometric data.
We suggest that the ionospheric free electron density can be modelled to good approximation by a Gaussian process restricted to a thick single layer, and show that under this assumption the differential total electron content must also be a Gaussian process.
We perform a comparison with a number of other widely successful Gaussian processes on simulated differential total electron contents over a wide range of experimental conditions, and find that, in all experimental conditions, our model is better able to represent observed data and generalise to unseen data.
The mean equivalent source shift imposed by our predictive errors are half as large as those of the best competitor model.
We find that it is possible to partially constrain the hyperparameters of the  ionosphere from sparse-and-noisy observed data.
Our model provides an alternative explanation for observed phase structure functions deviating from Kolmogorov's five-thirds turbulence, turnover at high baselines, and diffractive scale anisotropy.
We show that our model performs tomography of the free electron density both  implicitly and cheaply.
Moreover, we find that even a fast, low-resolution approximation of our model yields better results than the best alternative Gaussian process, implying that the geometric coupling between directions and antennae is a powerful prior that should not be ignored.
}
%We provide a solid mathematical basis for their use in modelling differential total electron content ($\Delta\mathrm{TEC}$) by proving that, under various conditions, GPs describe the correct stochastic nature of this observable.
%We fully develop a practical framework, elucidating quantitatively the statistical relations between the observable ($\Delta\mathrm{TEC}$) and the ionosphere's free electron density. We validate our results on simulated data.


   \maketitle

%________________________________________________________________

\section{Introduction}
Since the dawn of low-frequency radio astronomy, the ionosphere has been a confounding factor in the interpretation of radio data.
This is because the ionosphere has a spatially and temporally varying refractive index, which perturbs the radio-frequency radiation that passes through it. This effect becomes more severe at lower frequencies; see \citep[e.g.][]{degasperin2018}.
The functional relation between the sky brightness distribution -- the image -- and interferometric observables -- the visibilities -- is given by the radio interferometry measurement equation \citep[RIME;][]{1996A&AS..117..137H}, which models the propagation of radiation along geodesics from source to observer as an ordered set of linear transformations \citep{1941JOSA...31..488J}.

A mild ionosphere will act as a weak-scattering layer resulting in a perturbed inferred sky brightness distribution, analogous to the phenomenon of seeing in optical astronomy \citep{1969OptCo...1..153W}.
%That is, the average of many images with  changing ionospheric distortions results in a poor approximation of the true ionosphere-free sky.
Furthermore, the perturbation of a geodesic coming from a bright source will deteriorate the image quality far more than geodesics coming from faint sources.
Therefore, the image-domain effects of the ionosphere can be dependent on the distribution of bright sources on the celestial sphere, that is they can be heteroscedastic.
This severely impacts experiments which require sensitivity to faint structures in radio images.
Such studies include the search for the epoch of reionisation \citep[e.g.][]{2017ApJ...838...65P}, probes of the morphology of extended galaxy clusters \citep[e.g.][]{2019SSRv..215...16V}, efforts to detect the synchrotron cosmic web \citep[e.g.][]{Vernstrom2017}, and analyses of weak gravitational lensing in the radio domain \citep[e.g.][]{2016MNRAS.463.3674H}. Importantly, these studies were among the motivations for building the next generation of low-frequency radio telescopes like the Low Frequency Array (LOFAR), Murchison Widefield Array (MWA), and the future Square Kilometre Array (SKA).
Therefore, it is of great relevance to properly calibrate the ionosphere.
%It is important to note that such studies were among the motivations for 

% Martijn: you often mention *interferometric* visibilities - is the adjective really a valuable addition? Are there non-interferometric visibilities?
Efforts to calibrate interferometric visibilities have evolved over the years from single-direction, narrow-band, narrow-field-of-view techniques \citep{1973IEEEP..61.1192C}, to more advanced multi-directional, wide-band, wide-field methods \citep[e.g.][]{2011MNRAS.414.1656K,2016ApJS..223....2V, 2018A&A...611A..87T}.
The principle underlying these calibration schemes is that if you start with a rough initial model of the true sky brightness distribution, then you can calibrate against this model and generate an improved sky brightness model. One can then repeat this process for iterative improvement.
% calibrate based on image the sky and calibrate based on the revised image of the sky.
\review{Among the direction-dependent calibration techniques the most relevant for this paper is facet-based calibration, which applies the single-direction method to piece-wise independent patches of sky called facets.}
This scheme is possible if there are enough compact bright sources -- calibrators -- and if sufficient computational resources are available.
Ultimately, there are a finite number of calibrators in a field of view and additional techniques must be considered to calibrate all the geodesics involved in the RIME.
We note that there are other schemes for ionosphere calibration that do not apply the facet-based approach, such as image domain warping \citep{2017MNRAS.464.1146H}.

There are two different approaches for calibrating all geodesics involved in the RIME.
The first approach is to model the interferometric visibilities from first principles and then solve the joint calibration-and-imaging inversion problem.
This perspective is the most fundamental; however, applications \citep[e.g.][]{2016cvpr.conf..913B} of this type are very rare and often restricted to small data volumes due to exploding computational complexity.
However, we argue that investing research capital -- in small teams to minimize risk -- could be fruitful and disrupt the status quo \citep{smallteamsdisrupt}.
The second approach is to treat the piece-wise independent calibration solutions as data and predict calibration solutions for missing geodesics \citep[e.g.][]{2009A&A...501.1185I, 2016ApJS..223....2V, 2018A&A...611A..87T}.
In this paper, we consider an inference problem of the second kind.

In order to perform inference for the calibration along missing geodesics, a prior must be placed on the model.
One often-used prior is that the Jones operators are constant over some solution interval.
For example, in facet-based calibration the implicit prior is that two geodesics passing through the same facet and originating from the same antenna have the same calibration -- which can be thought of a nearest-neighbour interpolation.
\review{One often-neglected prior is the 3D correlation structure of the refractive index of the ionosphere.}
An intuitive motivation for considering this type of prior is as follows: The ionosphere has some intrinsic 3D correlation structure, and since cosmic radio emission propagates as spatially coherent waves. It follows that the correlation structure of the ionosphere should be present in ground-based measurements of the electric field correlation -- the visibilities.
The scope of this paper is therefore to build the mathematical prior corresponding to the above intuition.

We arrange this paper by first reviewing some properties of the ionosphere and its relation to interferometric visibilities via differential total electron content in Section~\ref{sec:ionosphere_intro}.
% Next, we define the ray integral (RI) and the related \textit{differenced} ray integral (DRI).
In Section~\ref{sec:model}, we then introduce a flexible model for the free electron density based on a Gaussian process restricted to a layer.
We derive the general relation between the probability measure for free electron density and differential total electron content, and use this to form a strong prior for differential total electron content along missing geodesics.
In Section~\ref{sec:method} we describe a numerical experiment wherein we test our model against other widely successful Gaussian-process models readily available in the literature. 
In Section~\ref{sec:resultA} we show that our prior outperforms the other widely successful priors in all noise regimes and levels of data sparsity.
Furthermore, we show that we are able to hierarchically learn the prior from data.
In Section~\ref{sec:discussion} we provide a justification for the assumptions of the model, and show the equivalence with tomographic inference.

\section{Ionospheric effects on interferometric visibilities}
\label{sec:ionosphere_intro}
The telluric ionosphere is formed by the geomagnetic field and a turbulent low-density plasma of various ion species, with bulk flows driven by extreme ultraviolet solar radiation \citep{1995isp..book.....K}.
Spatial irregularities in the free electron density (FED) $n_e$ and magnetic field $\vec{B}$ of the ionosphere give rise to a variable refractive index $n$, described by the Appleton-Hartree equation \citep{0741-3335-49-2-B01} -- here given in a Taylor series expansion to order $O(\nu^{-5})$:
\begin{align}
n(\x)\approx& 1 - \frac{\nu_p^2(\x)}{2\nu^2} \pm \frac{\nu_H(\x)\nu_p^2(\x)}{2\nu^3} - \frac{\nu_p^4(\x) - 4\nu_H^2(\x)\nu_p^2(\x)}{8\nu^4}.\label{eq:ah}
%\approx&1 - \frac{\nu_p^2(\x)}{2\nu^2} \pm \frac{\nu_H(\x)\nu_p^2(\x)}{2\nu^3}\cos\theta - \frac{\nu_p^4(\x)}{8\nu^4} - \frac{\nu_H^2(\x)\nu_p^2(\x)}{2\nu^4}\cos^2\theta.\label{eq:ah}
\end{align}
Here $\nu_p(\x) = \left(\frac{n_e(\x) q^2}{4\pi^2\epsilon_0 m}\right)^{1/2}$ is the plasma frequency, $\nu_H(\x) = \frac{B(\x) q}{2\pi m }$ is the gyro frequency,  $\nu$ is the frequency of radiation,
$q$ is the elementary charge, $\epsilon_0$ is the vacuum permittivity, and $m$ is the effective electron mass.
This form of the Appleton-Hartree equation assumes that the ionospheric plasma is cold and collisionless, that the magnetic field is parallel to the radiation wavevector, and that $\nu\gg \max\{\nu_p, \nu_H\}$.
The plus symbol corresponds to the left-handed circularly polarised mode of propagation, and the minus symbol corresponds to the right-handed equivalent. 
Going forward, we will only consider up to second-order effects, and therefore neglect all effects of polarisation in forthcoming analyses.

In the regime where refractive index variation over one wavelength is small, we can ignore diffraction and interference, or equivalently think about \emph{wave} propagation as \emph{ray} propagation \citep[e.g.][]{2010ApJ...718..963K}.
This approximation is known as the Jeffreys-Wentzel–Kramers–Brillouin approximation \citep{PLMS:PLMS0428}, which is equivalent to treating this as a scattering problem, and assuming that the scattered wave amplitude is much smaller than the incident wave amplitude -- the weak scattering limit \citep[e.g.][]{jres.066D.062,1969OptCo...1..153W}.
Light passing through a varying refractive index $n$ will accumulate a wavefront phase proportional to the path length of the geodesic traversed.
Let $\mathcal{R}_{\x}^{\hat{\vec{k}}}$ be a functional of $n$, so that the geodesic $\mathcal{R}_{\x}^{\hat{\vec{k}}}[n]: [0, \infty) \to \mathbb{R}^3$ maps from some parameter $s$ to points along it. 
The geodesic connects an Earth-based spatial location $\vec{x}$ to a direction on the celestial sphere, indicated by unit vector $\hat{\vec{k}}$.
% (We ignore pathological cases in which the pair $\left(\x, \hat{\vec{k}}\right)$ does not specify a unique geodesic $\mathcal{R}_{\x}^{\hat{\vec{k}}}[n]$.)
The accumulated wavefront phase along the path is then given by
\begin{align}
\phi_{\x}^{\hat{\vec{k}}}=& \frac{2\pi\nu}{c}\int_0^\infty n\left(\mathcal{R}_{\x}^{\hat{\vec{k}}}[n](s)\right) - 1\ \mathrm{d}s,\label{eq:fermat}
\end{align}
where $c$ is the speed of light \textit{in vacuo}. 
Hamilton's principle of least-action states that geodesics are defined by paths that extremise the total variation of Eq.~\ref{eq:fermat}.

By substituting Eq.~\ref{eq:ah} into Eq.~\ref{eq:fermat}, and by considering terms up to second order in $\nu^{-1}$ only, we find that the phase deviation induced by the ionosphere is proportional to the integral of the FED along the geodesic, $\phi_{\x}^{\hat{\vec{k}}} \approx \frac{- q^2}{4\pi\epsilon_0 mc \nu}\tau_{\x}^{\hat{\vec{k}}}$, where,
\begin{align}
\tau_{\x}^{\hat{\vec{k}}} \triangleq& \int_0^\infty n_e\left(\mathcal{R}_{\x}^{\hat{\vec{k}}}[n](s)\right) \, \mathrm{d}s.\label{eq:TEC}
\end{align}
Equation~\ref{eq:TEC} defines the total electron content (TEC).

In radio interferometry, the RIME states that the visibilities, being a measure of coherence, are insensitive to unitary transformations of the electric field associated with an electromagnetic wave.
Thus, the phase deviation associated with a geodesic is a relative quantity, usually referenced to the phase deviation from another fixed parallel geodesic -- the origin of which is called the reference antenna.
% We abuse the word `parallel' here, because there is no well-defined notion of angle between two space-like separated geodesics.
% Martijn: I would say that *in general* there is no well-defined notion of angle between vectors in different tangent spaces - e.g. those corresponding to different points on the manifold. Why do we specifically add the nuance of *space-like separated* geodesics?
% However, if we assume there is no curvature and mean that if the geodesics were transported to each other -- e.g. with Schild's ladder -- then they would be parallel.
Going forward we use Latin subscripts to specify geodesics with origins at an antenna location; for example  $\mathcal{R}_i^{\hat{\vec{k}}}[n]$ is used as shorthand for $\mathcal{R}_{\x_i}^{\hat{\vec{k}}}[n]$.
Correspondingly, we introduce the notion of differential total electron content ($\Delta\mathrm{TEC}$),
% : the difference between the TEC corresponding to a geodesic ending at some antenna of interest, and the TEC corresponding to a geodesic ending at a reference antenna, where the geodesics are assumed to originate from the same sky direction.
% \textit{In concreto}, we define the $\Delta\mathrm{TEC}$ between geodesics ending at antennae $i$ and $j$ and originating from direction $\hat{\vec{k}}$ to be
\begin{align}
\tau_{ij}^{\hat{\vec{k}}} \triangleq \tau_{i}^{\hat{\vec{k}}}  - \tau_{j}^{\hat{\vec{k}}}, \label{eq:DTEC}
\end{align}
which is the TEC of $\mathcal{R}_i^{\hat{\vec{k}}}[n]$ relative to $\mathcal{R}_j^{\hat{\vec{k}}}[n]$.

\section{Probabilistic relation between FED and $\Delta\mathrm{TEC}$: Gaussian process layer model}
\label{sec:model}
In this section we derive the probability distribution of $\Delta\mathrm{TEC}$ given a specific probability distribution for FED. 
It helps to first introduce the concept of the ray integral (RI) and the corresponding differenced ray integral (DRI).
The RI is defined by the linear operator $G_i^{\hat{\vec{k}}}: \mathcal{V} \to \mathbb{R}$ mapping from the space of all scalar-valued functions over $\mathbb{R}^3$ to a scalar value according to,
\begin{align}
G_i^{\hat{\vec{k}}} f \triangleq \int_0^\infty f\left(\mathcal{R}_{i}^{\hat{\vec{k}}}[n](s)\right) \, \mathrm{d}s,
\end{align}
where $f \in \mathcal{V} = \left\{ g \mid \int_{\mathbb{R}^3} g^2(\x) d\x < \infty\right\}$.
% (As the scalar field $n$ that determines the geodesics remains the same throughout the current discussion, we choose not to include it in our notation of $G_i^{\hat{\vec{k}}}$.)
Thus, an RI simply integrates a scalar field along a geodesic.
The DRI $\Delta_{ij}^{\hat{\vec{k}}}: \mathcal{V} \to \mathbb{R}$ for a scalar field $f$ is straightforwardly defined by
\begin{align}
\Delta_{ij}^{\hat{\vec{k}}} f \triangleq \left(G_{i}^{\hat{\vec{k}}} - G_{j}^{\hat{\vec{k}}}\right)f.\label{eq:dri_a}
\end{align}
% (Again, we choose to suppress a mention of $n$.)
Both the RI and DRI are linear operators in the usual sense.
Using Eqs.~\ref{eq:TEC} up to \ref{eq:dri_a}, we see that
\begin{align}
\tau_{ij}^{\hat{\vec{k}}} = \Delta_{ij}^{\hat{\vec{k}}} n_e.\label{eq:tau_b}
\end{align}

Let us now specify that the FED is a Gaussian process (GP) restricted to (and indexed by) the set of spatial locations $\mathcal{X} = \left\{ \x \in \mathbb{R}^3 \mid (\x - \x_0)\cdot \hat{\bs{z}} \in \left[a - b/2, a+b/2\right]\right\}$. 
This defines a layer of thickness $b$ at height $a$ above some reference point $\x_0$ (see \textbf{Figure}~\ref{fig:geometry}). 
Within this layer the FED is realised from,
\begin{align}
n_e \sim \mathcal{N}[\mu, K],
\end{align}
where $\mu:\mathcal{X} \to \mathbb{R}_{>0}$ is the mean function, and $K:\mathcal{X}\times\mathcal{X} \to \mathbb{R}$ is the covariance kernel function.
In other words, the ionospheric FED is regarded to be a uncountable infinite set of random variables (RVs) indexed by spatial locations in $\mathcal{X}$, such that for any finite subset of such locations the corresponding FEDs have a multivariate normal distribution. 
% For example, for any pair of coordinates $\x,\y \in \mathcal{X}$, the associated FEDs form a bivariate Gaussian, with the first non-central moment $m_1$ and the second central moment $m_2$ of the joint distribution being,
% \begin{align}
%     m_1 =& \left[\mu\left(\x\right), \mu\left(\y\right)\right]^T\\
%     m_2 =& \left[\begin{matrix}
%     K\left(\x, \x\right) & K\left(\x, \y\right) \\
%     K\left(\y, \x\right) & K\left(\y, \y\right) \\
%     \end{matrix}\right].
% \end{align}
% %with the first and second non-central moments $m_1$ and $m_2$ being 
% %\begin{align}
% %\mathbb{E}\left[n_e(\x)\right] =& m(\x)\\
% %\mathbb{E}\left[n_e(\x)n_e(\y)\right] =& K(\x,\y) + m(\x)m(\y).
% %\end{align}
In order to extend the scalar field $n_e$ to all of $\mathbb{R}^3$, so that we may apply the operator in Eq.~\ref{eq:dri_a} to FED, we impose that for all $\x \in \mathbb{R}^3 \setminus \mathcal{X}: n_e(\x) = 0$.
\review{This simply means that we take electron density to be zero outside the layer, and makes $G_{i}^{\hat{\vec{k}}}$ well-defined.}
To further simplify the model, we assume that the mean FED in the layer is constant; that is, for all \review{$\x \in \mathcal{X}: \mu(\x) = \bar{n}_e$.}
\begin{figure}[h]
    \centering
\def\svgwidth{\columnwidth}
    \caption{Geometry of the toy model. The ionosphere is a layer of thickness $b$ at height $a$ above a reference location $\x_0$. In general, $\Delta\mathrm{TEC}$ is the TEC along one geodesic minus the TEC along another parallel geodesic. Usually, these geodesics are originating at antennae $i$ and $j$ (locations $\vec{x}_i$ and $\vec{x}_j$), and pointing in directions $\hat{\vec{k}}_1$ and $\hat{\vec{k}}_2$, respectively. One common choice is to have a fixed reference antenna for all $\Delta\mathrm{TEC}$ measurements. The corresponding zenith angles are $\phi_1$ and $\phi_2$.}
\label{fig:geometry}
\end{figure}

One immediate question that arises pertains to the reasoning behind using a GP to model the FED in the ionosphere.
% Currently, the true distribution of the FED is indeterminate. 
\review{Currently, there is no adequate probabilistic description of the ionosphere that is valid for all times and at the spatial scales that we require.}
The state-of-the-art characterisation of the ionosphere at the latitude and scales we are concerned with are measurements of the phase structure function, a second-order statistic \citep{mevius2016}.
It is well known that second-order statistics alone do not determine a distribution.
In general, all moments are required to characterise a distribution, with a determinancy criterion known as Carleman's condition.
Furthermore, the ionosphere is highly dynamic and displays a multitude of behaviours.
\citet{2017MNRAS.471.3974J} observed four distinct behaviours of the ionosphere above the MWA.
It is likely that there are innumerable states of the ionosphere.

Due to the \review{above issue}, it is not our intent to precisely model the ionosphere.
We rather seek to describe it with a flexible and powerful probabilistic framework.
Gaussian processes have several attractive properties, such as the fact that they are highly expressive, easy to interpret, and (in some cases) allow closed-form analytic integration over hypotheses \citep{Rasmussen:2005:GPM:1162254}.

However, a Gaussian distribution assigns a non-zero probability density to negative values, which is unphysical.
One might instead consider the FED to be a log-GP, $n_e\left(\x\right) = \bar{n}_e \exp{\rho\left(\x\right)}$, where the dimensionless quantity $\rho\left(\x\right)$ is a Gaussian process. 
In the limit $\rho\left(\x\right) \to 0$, we recover that $n_e$ is itself a GP.
This is equivalent to saying that the $\sigma_{n_e} / \bar{n}_e \ll 1$.
As explained in Section~\ref{sec:method}, we determine estimates of $\sigma_{n_e}$ and $\bar{n}_e$ by fitting our models to actual observed calibrator data, the International Reference Ionosphere (IRI), and observations taken from \cite{1995isp..book.....K}.
This places the ratio at $\sigma_{n_e} / \bar{n}_e \lesssim 0.06$,
 suggesting that if the FED can be accurately described with a log-GP, then to good approximation it can also be described with a GP.

We now impose that the geodesics are straight rays, a simplification valid in the weak-scattering limit considered here.
The geodesics therefore become $\mathcal{R}_{\x}^{\hat{\bs{k}}}[n](s) = \x + s\hat{\bs{k}}$.
In practice, strong scattering due to small-scale refractive index variations in the ionosphere is negligible at frequencies far above the plasma frequency when the ionosphere is well-behaved, which is about 90\% of the time \citep{2015MNRAS.453..925V}.
For frequencies $\lesssim 50$~MHz however, this simplification becomes problematic.
Under the straight-ray assumption, Equation~\ref{eq:tau_b} becomes
\begin{align}
\tau_{ij}^{\hat{\vec{k}}} 
%=& \int_0^\infty n_e(\x_i + s\hat{\bs{k}})\,ds - \int_0^\infty n_e(\x_j + s'\hat{\bs{k}})\,ds' \\
=& \int_{s_i^{\hat{\vec{k}}-}}^{s_i^{\hat{\vec{k}}+}} n_e(\x_i + s\hat{\bs{k}})\,ds - \int_{s_j^{\hat{\vec{k}}-}}^{s_j^{\hat{\vec{k}}+}} n_e(\x_j + s'\hat{\bs{k}})\,ds' .\label{eq:tau_c}
\end{align}
Here, the integration limits come from the extension of the FED to spatial locations outside the index-set $\mathcal{X}$, and are given by
\begin{align}
s_i^{\hat{\vec{k}}\pm} = \left(a \pm \frac{b}{2} - \left(\x_i - \x_0\right) \cdot \hat{\vec{z}}\right) \sec \phi,
\end{align}
 where $\sec\phi = (\hat{\vec{k}}\cdot\hat{\bs{z}})^{-1}$ denotes the secant of the zenith angle.
It is convenient to colocate the reference point $\x_0$ with one of the antenna locations, and then to also specify this antenna as the reference antenna, i.e. the origin of all reference geodesics.
When this choice is made, $\Delta\mathrm{TEC}$ becomes $\tau_{i0}^{\hat{\vec{k}}}$.

Equation~\ref{eq:tau_b} shows directly that if $n_e$ is a GP, then so is  $\Delta\mathrm{TEC}$. 
This can be understood by viewing the RI as the limit of a Riemann sum. 
We reiterate that every univariate marginal of a multivariate Gaussian is also Gaussian, and that every finite linear combination of Gaussian RVs is again Gaussian.
Taking the Riemann sum to the infinitesimal limit preserves this property.
Since the DRI is a linear combination of two RIs, the result follows \citep[e.g.][]{JIDLING2018141}.

The index-set for the $\Delta\mathrm{TEC}$ GP is the product space of all possible antenna locations and vectors on the unit 2-sphere, $\mathcal{S} = \left\{(\x,\hat{\bs{k}}) \mid \x \in \mathbb{R}^3, \hat{\bs{k}} \in \mathbb{S}^2\right\}$.
\review{This is analogous to saying that the coordinates of the $\Delta\mathrm{TEC}$ GP are a tuple of antenna location and calibration direction.}
Thus, given any $\y = (\x, \hat{\bs{k}}) \in \mathcal{S,}$ the $\Delta\mathrm{TEC}$ \review{is denoted by} $\tau_{\x0}^{\hat{\bs{k}}}$.
\review{Because $\Delta\mathrm{TEC}$ is a GP, its distribution is completely specified by its first two moments.}
% In general, the $P$-th moment of $\Delta\mathrm{TEC}$ is given by,
% \begin{align}
% m_P =& \mathbb{E}\left[\prod_p^P \tau_{i_p0}^{\hat{\vec{k}}_p}\right] \\
% =&
% \mathbb{E}\left[\prod_p^P(G_{i_p}^{\hat{\vec{k}}_p} n_e - G_{0}^{\hat{\vec{k}}_p} n_e)\right].\label{eq:moments_a}
% \end{align}

Since we assume a flat layer geometry, the intersections of two parallel rays with the ionosphere layer have equal lengths of $b \sec \phi$.
This results in the mean TEC of two parallel rays being equal, and thus the first moment of $\Delta\mathrm{TEC}$ is,
\begin{align} 
m_{\Delta\rm TEC}(\y) =& 0,
\end{align}
where $\y=(\x_i, \hat{\bs{k}}) \in \mathcal{S}$.
It is important to note that this is not a trivial result.
Indeed, a more realistic but slightly more complicated ionosphere layer model would assume the layer follows the curvature of the Earth. 
In this case, the intersections of two parallel rays with the ionosphere layer have unequal lengths, and the first moment of $\Delta\mathrm{TEC}$ would depend on the layer geometry and $\bar{n}_e$.
 
We now derive the second central moment between two $\Delta\mathrm{TEC}$ along two different geodesics, as visualised in \textbf{Figure}~\ref{fig:geometry}.
\begin{align}
K_{\Delta\rm TEC}(\y,\y') =&\mathbb{E}\left[\tau_{i0}^{\hat{\vec{k}}} \tau_{j0}^{\hat{\vec{k}}'}\right] \\
=& \mathbb{E}\left[(G_{i}^{\hat{\vec{k}}} n_e - G_{0}^{\hat{\vec{k}}} n_e)(G_{j}^{\hat{\vec{k}}'} n_e - G_{0}^{\hat{\vec{k}}'} n_e)\right]\\
=& I_{ij}^{\hat{\vec{k}}\hat{\vec{k}}'} + I_{00}^{\hat{\vec{k}}\hat{\vec{k}}'} - I_{i0}^{\hat{\vec{k}}\hat{\vec{k}}'} - I_{0j}^{\hat{\vec{k}}\hat{\vec{k}}'},\label{eq:kernel_a}
\end{align}
where $\y=(\x_i, \hat{\bs{k}}) \in \mathcal{S}$ and $\y'=(\x_j, \hat{\bs{k}}') \in \mathcal{S}$ and, 
\begin{align}
    I_{ij}^{\hat{\vec{k}}\hat{\vec{k}}'} 
    %=& \mathbb{E}\left[G_{i}^{\hat{\vec{k}}} (n_e-\bar{n_e}) G_{j}^{\hat{\vec{k}}'} (n_e-\bar{n_e}) \right]\\
    % =&  \int_{s_i^{\hat{\vec{k}}-}}^{s_i^{\hat{\vec{k}}+}} \int_{s_j^{\hat{\vec{k}}'-}}^{s_j^{\hat{\vec{k}}'+}} \mathbb{E}\left[ n_e(\x_i + s\hat{\bs{k}}) n_e(\x_j + s'\hat{\bs{k}}')\right]\,dsds'\\
    =& \int_{s_i^{\hat{\vec{k}}-}}^{s_i^{\hat{\vec{k}}+}} \int_{s_j^{\hat{\vec{k}}'-}}^{s_j^{\hat{\vec{k}}'+}} K\left( \x_i + s\hat{\bs{k}},\x_j + s'\hat{\bs{k}}'\right)\,\mathrm{d}s\mathrm{d}s'.\label{eq:I}
\end{align}
We now see that the GP for $\Delta\mathrm{TEC}$ is zero-mean with a kernel that depends on the kernel of the FED and layer geometry.
The layer geometry of the ionosphere enters through the integration limits of Eq.~\ref{eq:I}.
Most notably, the physical kernel is non-stationary even if the FED kernel is.
\review{Non-stationarity means that the $\Delta\mathrm{TEC}$ model is not statistically homogeneous, a fact that is well known since antennae near the reference antenna typically have small ionospheric phase corrections.}
We henceforth refer to Eq.~\ref{eq:kernel_a} as the physical kernel, or our kernel.

\paragraph{\textit{Related work}.}
% \label{sec:relatedwork}
% 
Modelling the ionosphere with a layer has been used in the past.
\citet{jres.066D.062} performed analysis of transverse spatial covariances of wavefronts \citep[e.g.][]{chernov1961,keller1964stochastic} passing through the ionosphere.
Their layer model was motivated by the observation of scintillation of radio waves from satellites \citep{1959JGR....64.2281Y}.
One of their results is a simplified variance function, which can be related to the phase structure functions in Section~\ref{sec:var_func}.
In \citet{tol2009}, a theoretical treatment of ionospheric calibration using a layered ionosphere with Kolmogorov turbulence is done.
More recently, \citet{2016PASA...33...31A} attempted to model a variable-height ionosphere layer above the MWA using GPS measurements for the purpose of modelling a TEC gradient; however unfortunately they concluded that the GPS station array of the MWA is not dense enough to constrain their model.

\section{Method}
\label{sec:method}

In order to investigate the efficacy of the physical kernel for the purpose of modelling $\Delta\mathrm{TEC}$ we devise a simulation-based experiment.
\review{Firstly, we define several observational setups covering a range of calibration pierce-point sparsity and calibration signal-to-noise ratios.
A high signal-to-noise-ratio calibration corresponds to better determination of $\Delta\mathrm{TEC}$ from gains in a real calibration program.
Secondly, we characterise two ionosphere varieties as introduced in Section~\ref{sec:model}. 
Each ionosphere variety is defined by its layer height and thickness, and GP parameters.
For each pair of observational setup and ionosphere variety we realise FED along each geodesic and numerically evaluate Eq.~\ref{eq:tau_b} thereby producing $\Delta\mathrm{TEC}$.
We then add an amount of white noise to $\Delta\mathrm{TEC}$ which mimics the uncertainty in a real calibration program with a given calibration signal-to-noise ratio.
Finally, we compare the performance of our kernel against several other common kernels used in machine learning on the problem of Gaussian process regression, known as Kriging.
In order to do this, we generate $\Delta\mathrm{TEC}$ for extra geodesics and place them in a held-out dataset.
This held-out dataset is used for validation of the predictive performance to new geodesics given the observed $\Delta\mathrm{TEC}$.
We refer to the other kernels, which we compare our kernel to, as the competitor kernels, and the models that they induce, as the competitor models.}

\subsection{Data generation}

For all simulations, we have chosen the core and remote station configuration of LOFAR \citep{2013A&A...556A...2V}, which is a state-of-the-art low-frequency radio array centred in the Netherlands and spread across Europe.
The core and remote stations of LOFAR are located within the Netherlands with maximal baselines of 70~km, and we term this array the Dutch LOFAR configuration.
We thinned out the array such that no antenna is within 150~m of another.
We made this cutoff to reduce the data size because nearby antennae add little new information and inevitably raise computational cost.
For example, antennae like CS001HBA0 and CS001HBA1 are so close that their joint inclusion was considered redundant.

% We select a directional design layout following the Fibonacci spiral, e.g. as depicted in \textbf{Figure}~\ref{fig:simDusk}.
% This results in a slight clustering of calibrators in the central region of the field of view, which mimics selecting calibrators from a primary beam uncorrected radio image.

We consider several different experimental conditions, with a particular choice denoted by $\eta$, under which we compare our model to competitors.
We consider five levels of pierce-point sparsity: $\{10, 20, 30, 40, 50\}$ directions per field of view \review{($12.6~\mathrm{deg}^2$)}.
\review{For a given choice of pierce-point sparsity we place twice as many directions along a Fibonacci spiral -- scaled to be contained within the field of view -- and randomly select half of the points to be in the observed dataset and the other half to be in the held-out dataset. 
The Fibonacci spiral is slightly overdense in the centre of the field of view, which mimics selecting bright calibrators from a primary-beam uncorrected radio source model.}
We consider a range of \review{calibration signal-to-noise ratios, which correspond to Gaussian uncertainties of $\Delta\mathrm{TEC}$ that would be inferred from antenna-based gains in a real calibration program.}
We therefore consider 11 uncertainty levels on a logarithmic scale from 0.1 to 10~mTECU.
A typical state-of-the-art Dutch LOFAR-HBA (high-band antennae) direction-dependent calibration is able to produce on the order of 30 calibration directions \citep{2019A&A...622A...1S}, based on the number of bright sources in the field of view, and produce $\Delta\mathrm{TEC}$ with an uncertainty of approximately 1~mTECU; these levels of sparsity and noise probe above and below nominal LOFAR\review{-HBA} observing conditions.

We define an ionosphere variety as an ionosphere layer model with a particular choice of height $a$, thickness $b$, mean electron density $\bar{n}_e$, and FED kernel $K_\mathrm{FED}$ with associated hyperparameters, namely length-scale and variance.
As \review{mentioned} in Section~\ref{sec:model}, \review{due to the innumerable states of the ionosphere} our intent is not to exactly simulate the ionosphere, \review{but rather to derive a flexible model.} 
% ().
Therefore, to illustrate the flexibility of our model, we have chosen to experiment with two \review{very different ionosphere varieties which we designate the \textit{dawn} and \textit{dusk} ionosphere varieties.}
These ionosphere varieties are summarised in \textbf{Table}~\ref{tab:varieties}.
In Section~\ref{sec:var_func} we show that these \review{ionosphere varieties} predict phase structure functions which \review{are indistinguishable from} real observations.
In order to select \review{the layer height and thickness} parameters \review{for the dawn and dusk varieties} we took height profiles from the International Reference Ionosphere \citep[IRI;][]{2008AdSpR..42..599B} model. 
\begin{table}
\caption{Summary of the parameters of the simulated ionospheres.}              
\label{tab:varieties}      
\centering                                      
\begin{tabular}{lcccccc}         
\hline\hline
Variety & $a$~(km) & $b$~(km) & $K_{\rm FED}$ & $\sigma_{n_e}$~($\mathrm{m}^{-3}$) & HPD~(km)\\
\hline                                   
dawn & 250 & 100 & M32 & $6\cdot10^9$ & 15\\      
dusk & 350 & 200 & EQ & $3\cdot10^9$ & 15\\
\hline
\end{tabular}
\end{table}

\review{In order to choose the FED GP kernels and hyperparameters we note that} it has been suggested that scintillation is more pronounced during mornings, due to increased FED variation \citep[e.g.][]{1983A&A...120..313S}; therefore we chose a rough FED kernel for our dawn simulation.
Roughness corresponds to how much spectral power is placed on the shorter length-scales, and also relates to \textcolor[rgb]{0.984314,0.00784314,0.027451}{\textcolor[rgb]{0,0,0}{how differentiable realisations from the process are}}; e.g. see \textbf{Figure}~\ref{fig:kernels}. 
\review{For the dawn ionosphere we choose} the Mat\'ern-3/2 (M32) kernel,
\begin{align}
    K_{\rm M32}(\x,\x') = \sigma_{n_e}^2 \left( 1 + \frac{\sqrt{3}}{l_\mathrm{M32}}|\x - \x'|\right)\exp\left[\frac{-\sqrt{3}}{l_\mathrm{M32}} |\x - \x'|\right],\label{eq:m32}
\end{align}
which \review{produces realisations that are} only once differentiable and therefore rough.
\review{For the dusk ionosphere we choose} the exponentiated quadratic (EQ) kernel,
\begin{align}
    K_{\rm EQ}(\x,\x') = \sigma_{n_e}^2 \exp\left[\frac{-|\x - \x'|^2}{2 l_\mathrm{EQ}^2} \right],\label{eq:rbf}
\end{align}
which \review{produces realisations that are} infinitely differentiable and smooth.

Both kernels have two hyperparameters, variance $\sigma_{n_e}^2$ and length-scale $l$.
\review{In order to estimate the FED variation, $\sigma_{n_e}$,} we used observations from \citet{1995isp..book.....K} that TEC measurements are typically on the order of 10~TECU, with variations of about 0.1~TECU.
\review{Following the observation that the dawn typically exhibits more scintillation} we choose a twice higher $\sigma_{n_e}$ for our dawn simulation.
In addition to the length-scale we consider the half-peak distance (HPD) $h$, which corresponds to the distance at which the kernel reaches half of its maximum.
This parameter has a consistent meaning across all monotonically decreasing isotropic kernels, whereas the meaning of $l$ depends on the kernel. 
It is related to $h$ by $h \approx 1.177 l_\mathrm{EQ}$ for the EQ and $h \approx 0.969 l_\mathrm{M32}$ for the M32 kernel. 
The length-scales were chosen by simulating a set of ionospheres with different length-scales and choosing the length-scale that resulted in $\Delta\mathrm{TEC}$ screens that are visually similar to typical Dutch LOFAR-HBA calibration data. For a given ionosphere variety,  We note that this requires a much higher relative precision in the absolute TEC calculations.
Due to computational limits, we only realise one simulation per experimental condition -- that is, we do not average over multiple realisations per experimental condition -- however given the large number of experimental conditions there is enough variation to robustly perform an analysis.

\subsection{Competitor models}
\begin{figure*}
    \centering
    \includegraphics[width=\textwidth]{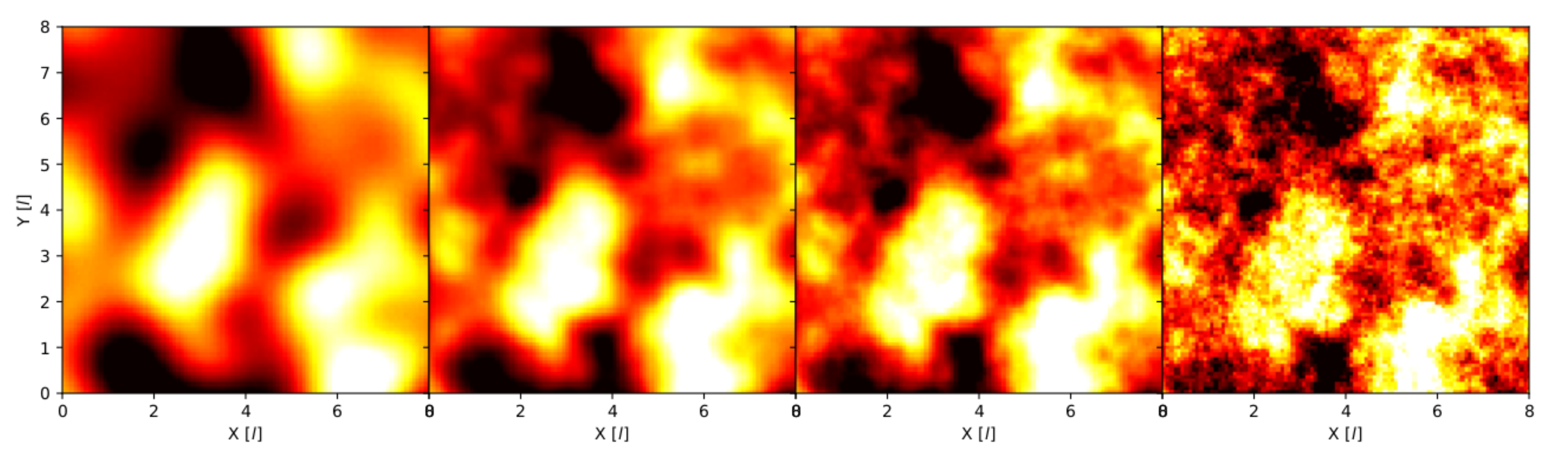}
    \caption{Example realisations from exponential quadratic, Mat\'ern-5/2, Mat\'ern-3/2, and Mat\'ern-1/2 kernels. The same HPD was used in all kernels, however the smoothness of the resulting process realisation is different for each.}
\label{fig:kernels}
\end{figure*}
\begin{figure}
    \centering
    \includegraphics[width=\columnwidth]{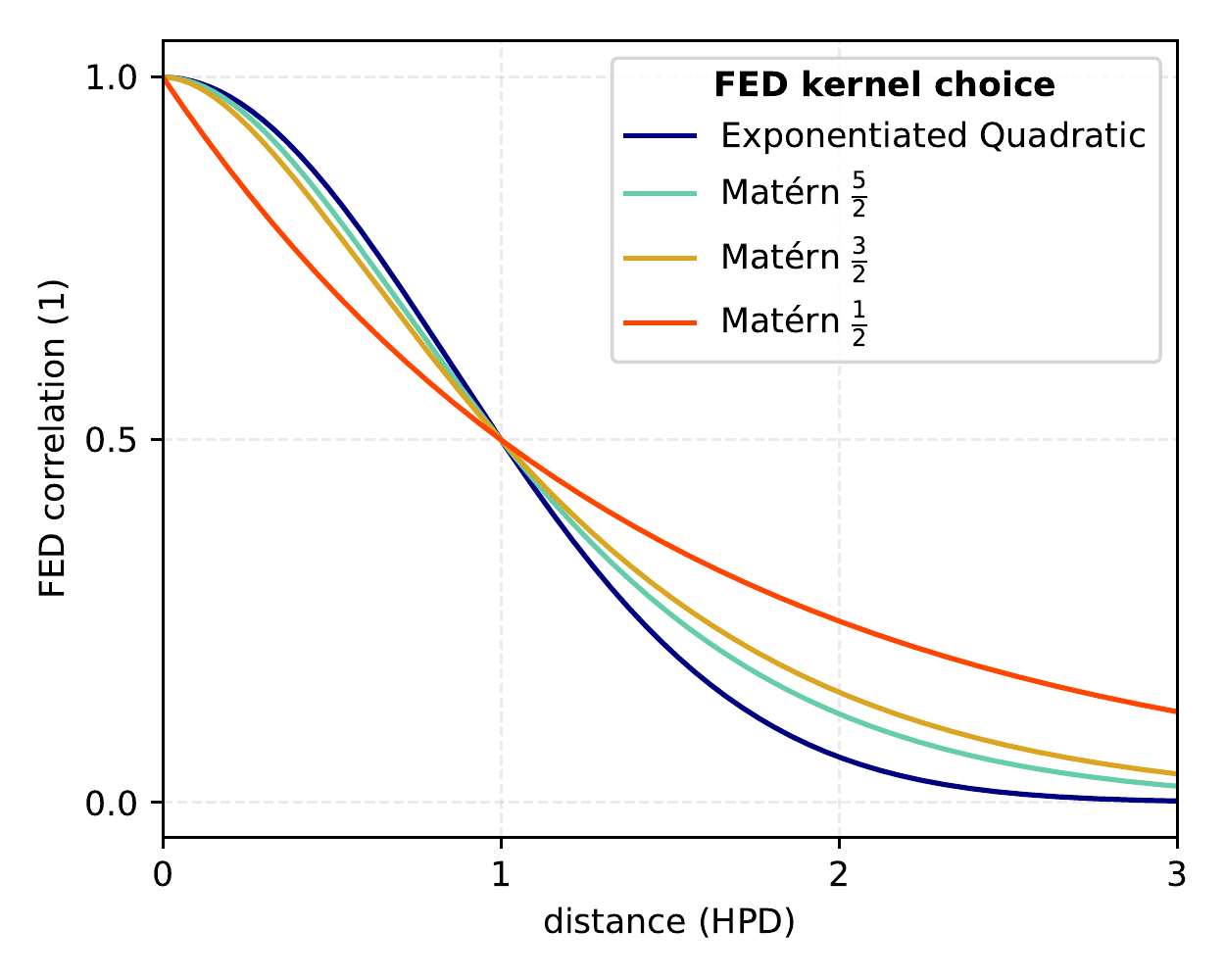}
    \caption{Shape of several kernels as a function of separation in units of the HPD of the kernel.}
    \label{fig:HPD}
\end{figure}

For the comparison with competitor models, we compare the physical kernel with: exponential quadratic (EQ), Mat\'ern-5/2 (M52), Mat\'ern-3/2 (M32), and Mat\'ern-1/2 (M12) \citep{Rasmussen:2005:GPM:1162254}.
The EQ and M32 kernels have already been introduced as FED kernels. 
The M52 and M12 are very similar except for having different roughness properties.
Each of these kernels results in a model that spatially smooths -- and therefore interpolates -- the observed data, but involves a different assumption on the underlying roughness of the function.
In order to use these kernels to model $\Delta\mathrm{TEC}$, we \review{give each subspace of $\mathcal{S}$ its own kernel and take the product}.
For example, if $K_C$ is the competitor kernel type, and \review{$(\x, \hat{\bs{k}}), (\x', \hat{\bs{k}}') \in \mathcal{S}$}, then we form the kernel \review{$K_C((\x, \hat{\bs{k}}), (\x', \hat{\bs{k}}')) = K^1_C(\x, \x')K^2_C(\hat{\bs{k}}, \hat{\bs{k}}')$} thereby giving each subspace of the index set, $\mathcal{S}$, its own kernel with associated hyperparameters.

\textbf{Figure}~\ref{fig:HPD} shows each kernel profile with the same HPD and \textbf{Figure}~\ref{fig:kernels} shows example realisations from the same kernels.
It can be visually verified that the M32 kernel has more small-scale variation than the EQ kernel, while maintaining similar large-scale correlation features.

% To minimise integration error, we calculated the accuracy of different numerical integration methods for various choices of $K_\mathrm{FED}$ and FED sampling density. 
% For a fixed FED sampling density, Simpson's rule was found to yield most precise results when $K_\mathrm{FED} = K_\mathrm{EQ}$ (see below), while the trapezoid rule worked best in case $K_\mathrm{FED} = K_\mathrm{M32}$ (idem). 
% The FED sampling densities were then chosen to guarantee $< 0.1\%$ integration error in case $K_\mathrm{FED} = K_\mathrm{EQ}$, and $< 1\%$ integration error in case $K_\mathrm{FED} = K_\mathrm{M32}$.

% a bit confusing and perhaps unnecessary:
%Whilst for any fixed HPD the EQ kernel provides higher FED correlations than the M32 kernel for distances smaller than the HPD, it provides lower FED correlations than the M32 kernel for distances larger than the HPD.

We note that evaluation of the physical kernel requires that a double integral be  performed, which can be done in several ways \citep[e.g.][]{2018arXiv181207319H}.
In our experiments we tried both explicit adaptive step-size Runge-Katta quadrature, and two-dimensional trapezoid quadrature.
We found via experimentation that we could simply use the trapezoid quadrature with each abscissa partitioned into four equal intervals without loss of effectiveness. However, we chose to use seven partitions.
We discuss this choice in Section~\ref{sec:low_acc}.

\subsection{Model comparison}

For model comparison, we investigate two key aspects of each model: the ability to accurately model observed $\Delta\mathrm{TEC}$, and the ability to accurately infer the held-out $\Delta\mathrm{TEC}$. 
In the language of the machine-learning community these are often referred to as minimising the data loss and the generalisation error, respectively.
We also investigate the ability to learn the hyperparameters of the physical kernel from sparse data.
Finding that the physical model accurately models both observed and held-out $\Delta\mathrm{TEC}$, while also being able to learn the hyper parameters, would be a positive outcome.

To measure how well a model represents the observed data, given a particular choice of kernel $K$ and
hyperparameters, we compute the log-probability of the observed (LPO) $\Delta\mathrm{TEC}$ data  -- Bayesian evidence -- which gives a measure of how well a GP fits the data with intrinsically penalised model complexity.
If we have data measured at $\X \in \mathcal{S}$ according to $\bs{\tau}_{\rm obs} = \bs{\tau}(\X) + \epsilon$ where $\epsilon \sim \mathcal{N}[0,\sigma^2]$ and $\bs{\tau}(\X)\sim\mathcal{N}[0,K(\X,\X)]$ then the LPO is,
\begin{align}
    \log P_{K}\left({\bs{\tau}_{\rm obs}} \right)
    %=& \log \int \mathcal{N}[\bs{\tau}_{\rm obs} \mid \bs{\tau}, \sigma^2 \bs{I}] \mathcal{N}[\bs{\tau} \mid 0, K_{\rm DTEC}(\X,\X)]\, d\bs{\tau}\\
    =& \log \mathcal{N}[0,  \bs{B}],\label{eq:LPO}
\end{align}
where $\bs{B} =  K(\X,\X) + \sigma^2 \bs{I}$.
To measure how well a model generalises to unseen data, given a particular choice of kernel $K$, we compute the conditional log-probability of held-out (LPH) data given the observed data.
That is, if we have a held-out dataset measured at $\X^* \in \mathcal{S}$ according to $\bs{\tau}_{\rm obs}^* = \bs{\tau}(\X^*) + \epsilon^*$ with $\epsilon^*\sim \mathcal{N}[0,\sigma^2]$ then the LPH conditional on observed $\bs{\tau}_{\rm obs}$ is,
\begin{align}
    \log P_K\left({\bs{\tau}_{\rm obs}^* \mid \bs{\tau}_{\rm obs}}\right)
    =& \log \mathcal{N}[K(\X^*,\X)\bs{B}^{-1}\bs{\tau}_{\rm obs}, \notag\\
    &\bs{B}^* - K(\X^*,\X)\bs{B}^{-1}K(\X,\X^*)]\label{eq:LPH}
\end{align}
where $\bs{B}^* = K(\X^*,\X^*) + \sigma^2\bs{I}$.
% In general, the held-out data can have a different observational variance $\sigma_*^2$ than the observed data, and in our case we are interested in the ability to predict noise-free held-out data, $\sigma_*^2 \to 0$, which we call the ground truth.
% This gives a measure of how well the kernel is able to infer missing $\Delta\mathrm{TEC}$ given some observed $\Delta\mathrm{TEC}$.
% For example, if radio astronomers were able to observe $\Delta\mathrm{TEC}$ along some geodesics and wanted to infer the $\Delta\mathrm{TEC}$ along extra geodesics, then this is the measure that would be most important to maximise.
% Higher values of LPO and LPH imply a better ability of a model, with a particular choice of kernel $K$, to represent the observed, and unseen data respectively. 

In order to make any claims of model superiority, we will define the following two figures of merit (FOMs),
\begin{align}
    \mathrm{\Delta LPO}_{\rm C}(\eta) \triangleq&  \frac{P_{\rm \Delta TEC} \left(\bs{\tau}_{\rm obs} \mid \eta \right)}{P_{\rm C} \left(\bs{\tau}_{\rm obs} \mid \eta \right)},\\
    \mathrm{\Delta LPH}_{\rm C}(\eta) \triangleq&  \frac{P_{\rm \Delta TEC} \left(\bs{\tau}_{\rm obs}^* \mid \bs{\tau}_{\rm obs}, \eta \right)}{P_{\rm C} \left(\bs{\tau}_{\rm obs}^* \mid \bs{\tau}_{\rm obs}, \eta \right)},
\end{align}
where $P_{\rm \Delta TEC}$ is the probability distribution using the physical kernel and $P_{\rm C}$ is the distribution using a competitor kernel.
The variable $\eta$ represents a particular choice of experimental conditions, for example pierce point sparsity and noise.

These FOMs specify how much more or less probable the physical kernel model is than a competitor for the given choice of experimental conditions, and are therefore useful interpretable numbers capable of discriminating between two models.
For example, a ${\mathrm{\Delta LPO}_{\rm C}(\eta)}$ value of 1 implies that for the given experimental conditions, $\eta$, both models have an equal probability of  representing the observed data, and a value of 1.5 would imply that the physical kernel representation is 50\% more probable than the competitor kernel.
We note that considering the ratio of marginal probabilities is the canonical way of model selection \citep{Rasmussen:2005:GPM:1162254}.
For a rule-of-thumb using these FOMs, we empirically visually find that models produce noticeably better predictions starting at around 1.10 (10\%).

For each choice of experimental conditions, $\eta,$ and kernel model, we first infer the maximum \textit{a posteriori} estimate of the hyperparameters of the kernel by maximising the marginal log-likelihood of the corresponding GP \citep{Rasmussen:2005:GPM:1162254}, which is equivalent to maximising the LPO of that model on the available observed dataset.
We maximise the marginal log-likelihood using the variable metric BFGS method, which uses a low-rank approximation to the Hessian to perform gradient-based convex optimisation \citep{bfgs}.
We use the GPFlow library \citep{GPflow2017}, which simplifies the algorithmic process considerably.
On top of this we perform optimisation from multiple random initialisations to avoid potential local minima.
For the physical kernel this corresponds to learning the layer height $a$ and thickness $b$, and FED kernel length-scale $l$, and variance $\sigma_{n_e}^2$, and for the competitor kernels this corresponds to learning a variance and the length-scales for each subspace.

\section{Results}
\label{sec:results}
\label{sec:resultA}

\begin{figure*}
    \centering
    % \begin{subfigure}{.93\textwidth}
    % \includegraphics[width=\textwidth]{simOverviewAntennaScreens_set24_DTEC_points_2Sigma.pdf}
    % \end{subfigure}
    % \begin{subfigure}{.93\textwidth}
    \includegraphics[width=\textwidth]{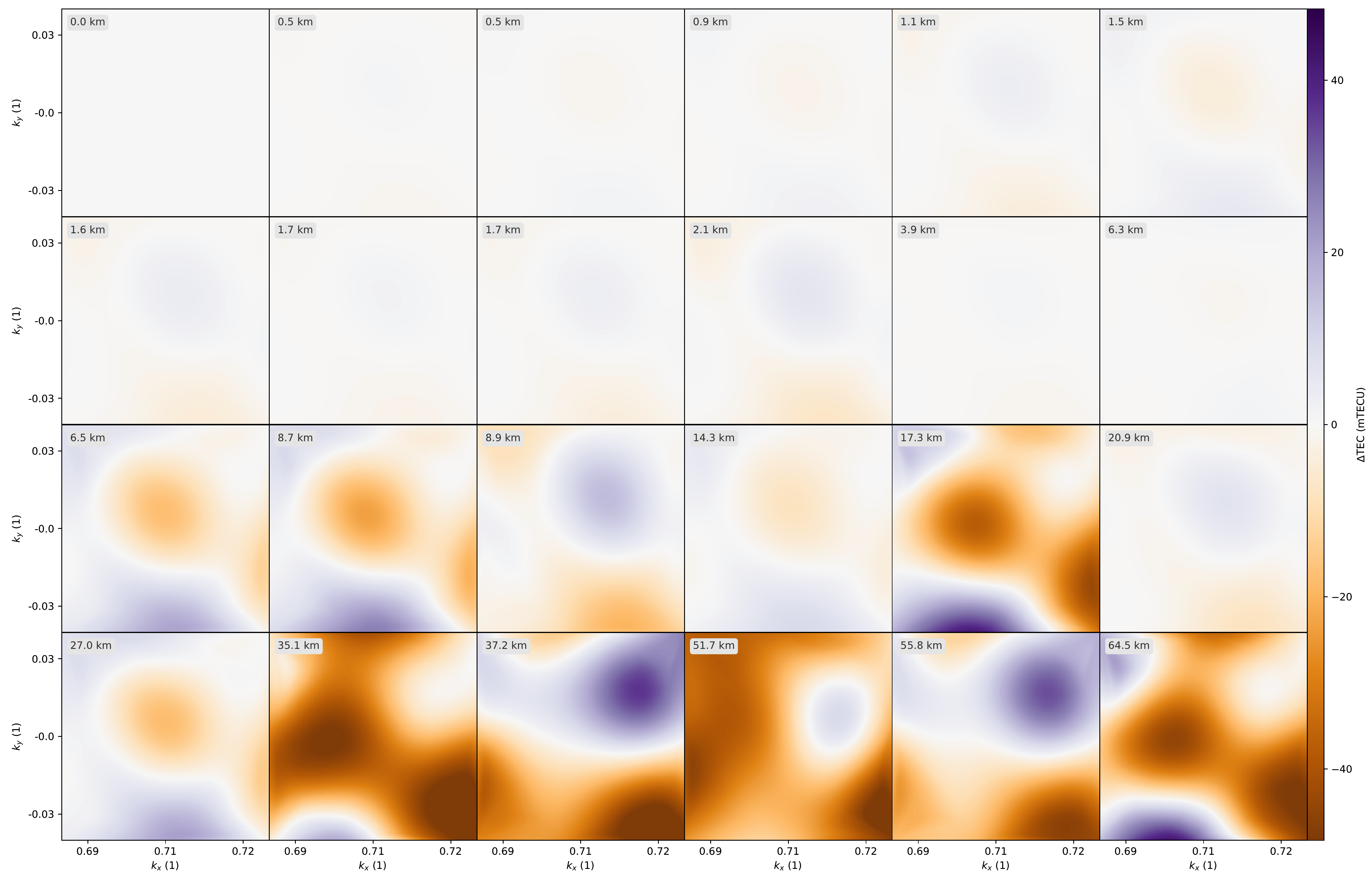}
    % \end{subfigure}
    \caption{Example of antenna-based $\Delta\mathrm{TEC}$ screens from the dusk ionosphere simulation. Each plot shows the simulated ground truth (noise-free) $\Delta\mathrm{TEC}$ for each geodesic originating from that station with axes given in direction components $k_x$ and $k_y$. The inset label gives how far the antenna is from the reference antenna. Antennae further from the reference antenna tend to have a larger magnitude $\Delta\mathrm{TEC}$ as expected. 
    Each plot box bounds a circular $12.6~\mathrm{deg}^2$ field of view. }
    \label{fig:simDusk}
\end{figure*}
In \textbf{Table}~\ref{tab:hp_learn} we report the average and standard deviation, over all experimental conditions, of the difference between the learned physical hyperparameters and the true hyperparameters, which we term the discrepancy.
 The optimisation converged in all cases.
We observe that for both ionosphere varieties the discrepancy of $a$ is on the order of a $\sim 10~\mathrm{km}$, or a few percent, implying that $a$ can be learned from data.
The discrepancy of HPD, is on the order of $1~\mathrm{km}$, or around $10\%$, implying the spectral shape information of the FED can be constrained from data.
We observe that the discrepancy of layer thickness, $b$, is large and on the order of $50\%$.
One reason for this is because Eq.~\ref{eq:I} will scale to first order with $b$ -- which is degenerate with the function of $\sigma_{n_e}$ -- and the only way to break the degeneracy is to have enough variation in the secant of the zenith angle.
In a sparse and noisy observation of $\Delta\mathrm{TEC}$, the secant variation is poor and it is expected that this degeneracy exists.
Therefore we also show the product $b \sigma_{n_e}$, and we see that this compound value discrepancy is smaller by approximately $35\%$.
% Another source of hyperparameter reconstruction discrepancy comes from the numerical integration required to calculate the physical kernel.
% As noted, we chose to discretise each line integral abscissa into 4 partitions, which results in a relative kernel error of about $10\%$ in the kernel calculation.
% It is possible that the effective meaning of the hyperparameters then deviates from their physical meanings due to the numerical approximation of the actual kernel.
% The physical kernel can be used to study the ionosphere via inference of the hyperparameters, however then more partitions should be used.
\begin{table}
\caption{Average and standard deviation, over all experimental conditions, of the difference between the learned physical hyperparameters and the true hyperparameters.}              
\label{tab:hp_learn}      
\centering                                      
\begin{tabular}{lcccc}         
\hline\hline
Variety & $a$ & $b$ & HPD& $b \sigma_{n_e}$\\
 {} & (km) & (km) & (km) & $(10^{11} \mathrm{km}\,\mathrm{m}^{-3}$)\\
\hline
% dawn & $10$ & $48$ &$4$ & $1.9$\\
% {} & $4\%$ & $48\%$ &$24\%$ & $32\%$\\
% dusk & $16$ & $82$ &$1$ & $2.2$\\
% {} & $5\%$ & $41\%$ & $10\%$ & $36\%$\\
dawn & $10 \pm 10$ & $48 \pm 18$ &$4 \pm 3$ & $1.9 \pm 1.2$\\
% {} & $(4 \pm 4)\%$ & $(48 \pm 18)\%$ &$(24 \pm 18)\%$ & $(32 \pm 20)\%$\\
dusk & $16 \pm 9$ & $82 \pm 20$ &$1 \pm 0.5$ & $2.2 \pm 0.3$\\
% {} & $(5 \pm 3)\%$ & $(41 \pm 10)\%$ & $(10 \pm 3)\%$ & $(36 \pm 5)\%$\\
\hline
\end{tabular}
\end{table}

% In \textbf{Figure}~\ref{fig:resultAa} the upper image plots show the $\mathrm{\Delta LPO}_{\rm C}(\eta)$ for each competitor kernel, and the lower image plots show the $\mathrm{\Delta LPH}_{\rm C}(\eta)$ for each competitor kernel.
% The image axes are the number of directions in the observed dataset, and observation noise.
% The left four columns are the results for the dawn ionosphere, and the right four are for the dusk ionosphere.
% We observe that the $\mathrm{\Delta LPO}_{\rm C}(\eta)$ and $\mathrm{\Delta LPH}_{\rm C}(\eta)$ is positive, for both the dawn and dusk ionosphere varieties, and for all combinations of observation noise and number of directions.
% This implies that the physical kernel produces a more probable model for all simulation parameters we chose.

In \textbf{Table}~\ref{tab:log_ratios} we summarise the performance of the physical kernel against each competitor kernel.
We display the mean of ${\mathrm{\Delta LPO}_{\rm C}(\eta)}$, and ${\mathrm{\Delta LPH}_{\rm C}(\eta)}$ over all experimental conditions, as well as their values at the nominal experimental conditions of 30 directions per $12.6~\mathrm{deg}^2$, and $\Delta\mathrm{TEC}$ noise of 1~mTECU, which is indicated with $\eta_{\rm nom}$.
We use bold font in \textbf{Table}~\ref{tab:log_ratios} to indicate the best competitor model.
% As discussed above, the interpretation of the number corresponds to how much more probable the physical kernel model is that the competitor model.

We first consider the ability of each model to represent the observed data.
For the dawn ionosphere, the M52 competitor kernel has the best (lowest) ${\langle\mathrm{\Delta LPO}_{\rm C}\rangle_\eta} = 1.55$  and $\Delta \mathrm{LPO}_{\rm C}^{\eta_{\rm nom}} = 1.46$, implying that the M52 kernel model is 55\% and 46\% less probable than the physical kernel model on average over all experimental conditions, and at nominal conditions, respectively.
We note that the M32 kernel produced similar results.
For the dusk ionosphere, the EQ kernel model is likewise the best among all competitors, being only 73\% and 54\% less probable than the physical kernel model on average over all experimental conditions, and at nominal conditions, respectively.
In all experimental conditions, the physical model provides a significantly more probable explanation of the observed data.

We now consider the ability of each model to infer the held-out data.
For the dawn ionosphere, the M52 competitor kernel has the best (lowest) ${\langle\mathrm{\Delta LPH}_{\rm C}\rangle_\eta} = 1.49$  and ${\mathrm{\Delta LPO}_{\rm C}^{\eta_{\rm nom}}} = 1.31$, implying that the M52 kernel prediction is 49\% and 31\% less probable than the physical kernel model on average over all experimental conditions, and at nominal conditions, respectively.
We note that the M32 kernel produced similar results.
For the dusk ionosphere, the EQ kernel model is likewise the best among all competitors, with predictions only 16\% and 12\% less probable than the physical kernel model on average over all experimental conditions, and at nominal conditions, respectively.
In the case of the rougher dawn ionosphere, the physical model provides a significantly more probable prediction of the held-out data in all experimental conditions. However, for the smoother dusk ionosphere at nominal conditions, the physical model is only 12\% more probable than the EQ kernel model, which is not very significant.

% \begin{table*}
% \caption{Log ratios of probabilities between the $\Delta\mathrm{TEC}$ and competitor kernels}      
% \label{tab:log_ratios}      
% \centering                                      
% \begin{tabular}{lcccc}         
% \hline\hline
% {}& $\langle\mathrm{\Delta LPO}_{\rm C}\rangle_\eta$ & $\mathrm{\Delta LPO}_{\rm C}(\eta_{\rm nom})$  & $\langle\mathrm{\Delta LPH}_{\rm C}\rangle_\eta$ & $\mathrm{\Delta LPH}_{\rm C}(\eta_{\rm nom})$\\
% \hline
% \multicolumn{5}{c}{dawn}\\
% M12 & 0.62 (1.86) & 0.58 (1.79) & 0.60 (1.82) & 0.48 (1.61)\\
% M32 & \textbf{0.44 (1.56)} & \textbf{0.40 (1.49)} & \textbf{0.40 (1.50)} & \textbf{0.29 (1.33)}\\
% M52 & \textbf{0.44 (1.55)} & \textbf{0.38 (1.46)} & \textbf{0.40 (1.49)} & \textbf{0.27 (1.31)}\\
% EQ & 0.49 (1.63) & 0.39 (1.48) & 0.61 (1.84) & 0.30 (1.35)\\
% \multicolumn{5}{c}{dusk}\\
% M12 & 1.00 (2.72) & 0.78 (2.19) & 0.81 (2.24) & 0.55 (1.73)\\
% M32 & 0.67 (1.96) & 0.52 (1.69) & 0.40 (1.50) & 0.25 (1.29)\\
% M52 & 0.60 (1.82) & 0.47 (1.60) & 0.28 (1.33) & 0.18 (1.20)\\
% EQ & \textbf{0.55 (1.73)} & \textbf{0.43 (1.54)} & \textbf{0.15 (1.16)} & \textbf{0.12 (1.12)}\\
% \hline
% \end{tabular}
% \end{table*}

\begin{table}
\caption{Shows the probability ratio FOMs (see text) averaged over experimental conditions and at nominal conditions. Larger values indicate that the physical model is more probable. Bold face indicates the best performing competitor model (lower number).}      
\label{tab:log_ratios}      
\centering                                      
\begin{tabular}{lcccc}         
\hline\hline
{}& ${\langle\mathrm{\Delta LPO}_{\rm C}\rangle_\eta}$ & $\mathrm{\Delta LPO}_{\rm C}^{\eta_{\rm nom}}$  & ${\langle\mathrm{\Delta LPH}_{\rm C}\rangle_\eta}$ & $\mathrm{\Delta LPH}_{\rm C}^{\eta_{\rm nom}}$\\
\hline
\multicolumn{5}{c}{dawn}\\
M12 & 1.86 & 1.79 & 1.82 & 1.61\\
M32 & {1.56} &  {1.49} &  {1.50} &  {1.33}\\
M52 & \textbf{1.55} & \textbf{1.46} & \textbf{1.49} & \textbf{1.31}\\
EQ & 1.63 & 1.48 & 1.84 & 1.35\\
\multicolumn{5}{c}{dusk}\\
M12 & 2.72 & 2.19 & 2.24 & 1.73\\
M32 & 1.96 & 1.69 & 1.50 & 1.29\\
M52 & 1.82 & 1.60 & 1.33 & 1.20\\
EQ & \textbf{1.73} & \textbf{1.54} & \textbf{1.16} & \textbf{1.12}\\
\hline
\end{tabular}
\end{table}

\begin{figure*}[h]
    \centering
    \includegraphics[clip, trim=1.5cm 0cm 0cm 1.5cm, width=\textwidth]{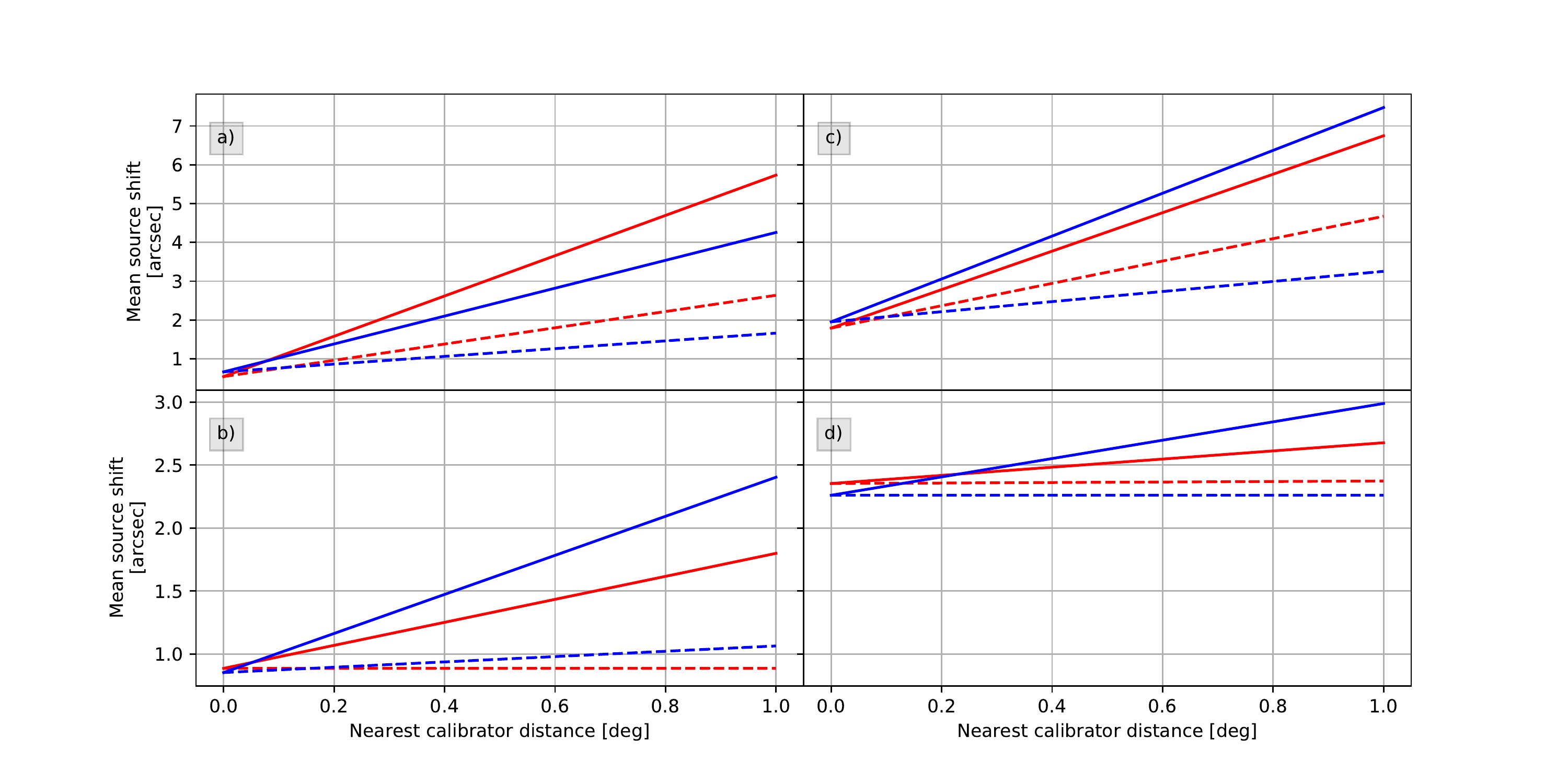}
    \caption{Mean equivalent source shift as a function of angular distance from the nearest calibrator caused by inference errors from the ground truth for \textbf{a)} remote stations (RS; $>3~\mathrm{km}$ from the reference antenna) at nominal conditions (30 calibrators for $12.6\mathrm{deg}^{2}$ and $1~\mathrm{mTECU}$ noise), \textbf{b)} core stations (CS; $<2~\mathrm{km}$) at nominal conditions, \textbf{c)} RS with sparse-and-noisy conditions (10 calibrators for $12.6\mathrm{deg}^{2}$ and $2.6~\mathrm{mTECU}$ noise), and \textbf{d)} CS with sparse-and-noisy conditions. The \textit{solid line styles} are the best competitor models (see text), the \textit{dashed line styles} are the physical model. The red lines are dawn ionospheres, and the blue lines are dusk ionospheres.}
\label{fig:resultAb}
\end{figure*}

\begin{figure*}[h]
\centering
\includegraphics[width=\textwidth]{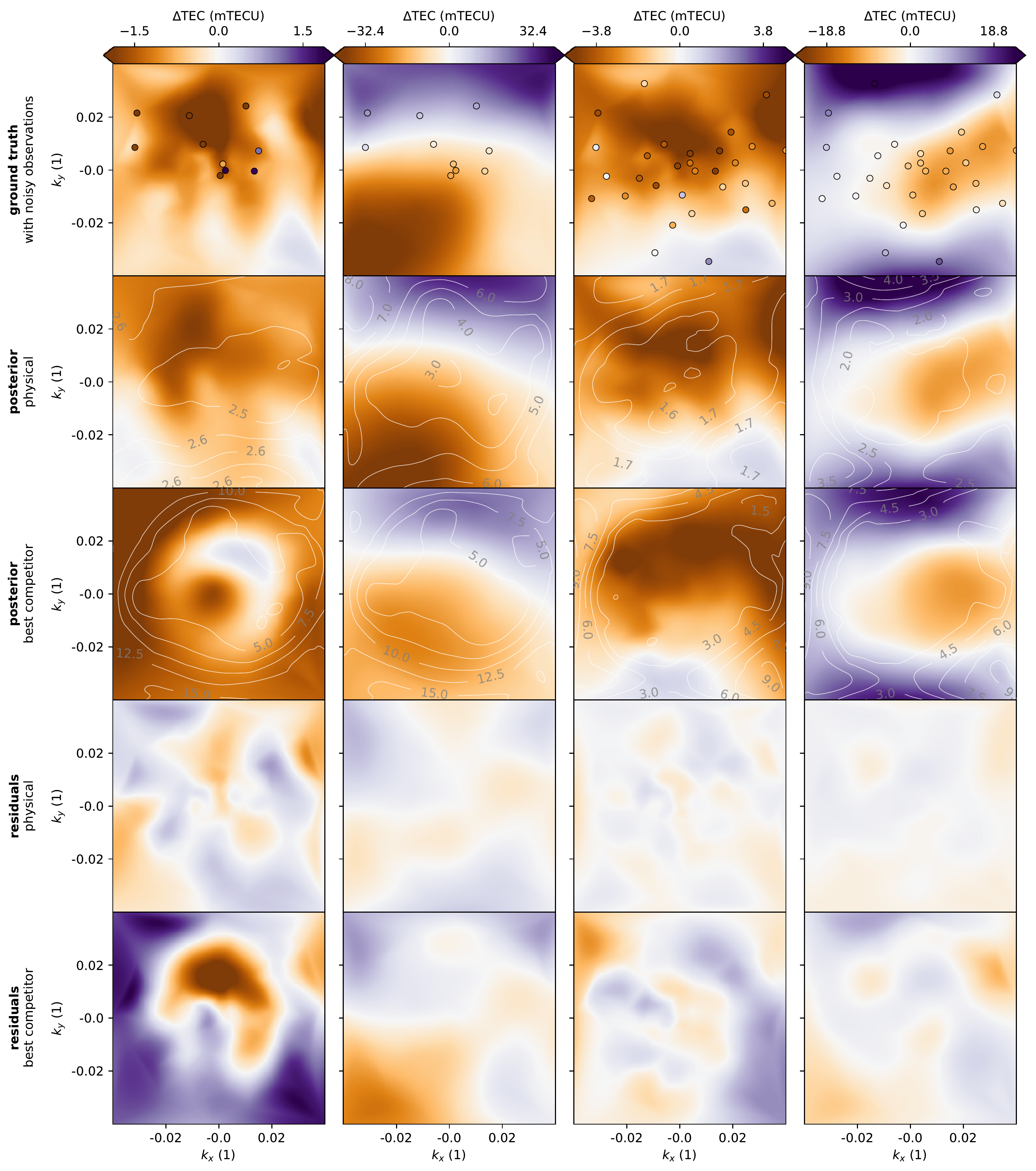}
\caption{Example visual comparison of the predictive performance of our physical model with that of the best competitor model for the dawn ionosphere.
\textit{First row} -- ground truth $\Delta\mathrm{TEC}$ overlaid on noisy draws from the ground truth which are the observations; \textit{Second and third rows} -- posterior mean with uncertainty contours for the physical model and best competitor model respectively.
\textit{Fourth and fifth rows} -- residuals between posterior means and ground truth for the physical model and best competitor model respectively.
\textit{First two columns:} Results for experimental conditions, (10 directions, 2.5~mTECU noise), for a central antenna (near to reference antenna) and a remote station (far from reference antenna);
\textit{Last two columns:}  Results for experimental conditions, (30 directions, 1.6~mTECU noise), for a central antenna and a remote station.
}
\label{fig:posterior_screens}
\end{figure*}

\textbf{Figure}~\ref{fig:posterior_screens} shows a visual comparison of the predictive distributions of the physical and best competitor kernel for the dawn ionosphere, for nominal and sparse-and-noisy conditions, for a subset of antennae over the field of view. 
In the first row we show the ground truth and observed data.
In the second and third rows we plot the mean of the predictive distribution with uncertainty contours of the physical and best competitor models, respectively.
At nominal conditions, the  predictive
means of the best competitor and physical models both visually appear to follow the shape of the ground truth.
However, for the sparse-and-noisy condition, only the physical model predictive mean visually follows the shape of the ground truth.
The uncertainty contours of the physical model vary in height slowly over the field of view, and are on the order of 0.5--1~mTECU. 
The uncertainty contours for the physical model indicate that we can trust the predictions near the edges of the field of view.
In comparison, the uncertainty contours of the best competitor model steeply grow in regions without calibrators, and are on the order of 2--10~mTECU, indicating that only predictions in densely sampled regions should be trusted.

The last two rows show the residuals between the posterior means and the ground truth for the physical and best competitor models respectively.
From this we can see that even when the best-competitor predictive mean visually appears to follow the ground truth the residuals are larger in magnitude than those of the physical models.

In order to quantify the effect of the residuals, a $\Delta\mathrm{TEC}$ error, $\delta \tau$, can be conveniently represented by the equivalent source shift for a source at zenith on a baseline of $r$, 
\begin{align}
    \delta l \approx& \frac{q^2}{\epsilon_0 m_e \nu^2 r}\delta \tau \\
    \approx& 1.16\arcsec \left(\frac{r}{10\mathrm{km}}\right)^{-1}\left(\frac{\nu}{150\mathrm{MHz}}\right)^{-2}\left(\frac{\delta \tau}{\mathrm{mTECU}}\right).
\end{align}
\textbf{Figure}~\ref{fig:resultAb} shows the mean linear regression of the absolute equivalent source shift of the residuals for each point in the held-out data set, for nominal (left) and sparse-and-noisy (right) conditions, at 150~MHz on a baseline of 10~km, as a function of the nearest calibrator.
For visual clarity we have not plotted confidence intervals, however we note that for nominal conditions the $1\sigma$ confidence width is about $2\arcsec$ and for the sparse-and-noisy conditions it is about $4\arcsec$.
Because there are few nearest-calibrator distances exceeding 1~degree at nominal conditions, we only perform a linear regression out to 1~degree.

The upper row shows the source shift for the remote stations (RS) residuals, which are generally much larger than the source shifts for core stations (CS) in the bottom row, since the CS antennae are much closer to the reference antenna and have smaller $\Delta\mathrm{TEC}$ variance.
We observe that the physical model (dashed line styles) generally has a shallower slope than the best competitor model (solid line styles).
Indeed, for the CS antennae the physical model source shift is almost independent of distance from a calibrator.
The offset from zero at 0 degrees of separation comes from the fact that the predictive variance cannot be less than the variance of the observations; see the definition of $\bs{B}^*$ in Eq.~\ref{eq:LPH}.
At 1 degree of separation, the physical model mean equivalent source shift is approximately half of  that of the best competitor model.
At 0 degrees of separation, the mean source shift is the same for both models as expected.

\section{Discussion}
\label{sec:discussion}

\subsection{Model selection bias}

\review{Our derived model is a probabilistic model informed by the physics of the problem.
We use the same physical model to simulate the data.
Therefore it should perform better than any other general-purpose model.
The fact that we simulate from the same physical model as used to derive the probabilistic model does not detract from the efficacy of the proposed model to represent the data.
Indeed, it should be seen as a reason for preferring physics-based approaches when the physics are rightly known.
The Gaussian random field layer model for the ionosphere has been a useful prescription for the ionosphere for a long time \citep[e.g.][]{1959JGR....64.2281Y}.}

\review{One type of bias that should be addressed is the fact that we assume we know the FED kernel type of the ionosphere.
We do not show, for example, what happens when we assume the wrong FED kernel.
However, since we are able to converge on optimal hyper parameters for a given choice of FED kernel, we can therefore imagine performing model selection based on the values of the Bayesian evidence (LPO) for different candidate FED kernels.
Thus, we can assume that we could correctly select the right FED kernel in all the experimental conditions that we chose in this work.}

\subsection{Implicit tomography}
\label{sec:imp_tomo}
The results of Section~\ref{sec:results} indicate that the physical model provides a better explanation of $\Delta\mathrm{TEC}$ data than any of the competitor models.
One might ask how  it performs so well.
The approach we present is closely linked to tomography, where (possibly non-linear) projections of a physical field are inverted for a scalar field.
In a classical tomographic approach, the posterior distribution for the FED given observed $\Delta\mathrm{TEC}$ data would be inferred and then the predictive $\Delta\mathrm{TEC}$ would be calculated from the FED, marginalising over all possible FEDs,
\begin{align}
    \prob{\bs{\tau} \mid  \bs{\tau}_{\rm obs}} = \int_{\bs{n_e}} \prob{\bs{\tau} \mid \bs{n_e}} \prob{\bs{n_e} \mid \bs{\tau}_{\rm obs}}\, \mathrm{d}\bs{n_e},\label{eq:jointcond_a}
\end{align}
where $\bs{n_e} = \{n_e(\x) \mid \x \in \mathcal{X}\}$ is the set of FEDs over the entire index set $\mathcal{X}$, $\bs{\tau} = \{\bs{\tau}_{\x}^{\hat{\bs{k}}} \mid (\x, \hat{\bs{k}}) \in \mathcal{S}_{*} \subset \mathcal{S}\}$ is the $\Delta\mathrm{TEC}$ over some subset $\mathcal{S}_{*}$ of the index set $\mathcal{S}$, $\bs{\tau}_{\rm obs} = \{\bs{\tau}_{\x}^{\hat{\bs{k}}} + \epsilon \mid (\x, \hat{\bs{k}}) \in \mathcal{S}_{\rm obs} \subset \mathcal{S}\}$ is the observed $\Delta\mathrm{TEC}$ over a different subset $\mathcal{S}_{\rm obs}$ of $\mathcal{S,}$ and $\epsilon \sim \mathcal{N}[\bs{0}, \sigma^2\bs{I}]$.

In our model, the associated equation for $\prob{\bs{\tau} \mid  \bs{\tau}_{\rm obs}}$ is found by conditioning the joint distribution on the observed $\Delta\mathrm{TEC}$ and then marginalising out FED,
\begin{align}
    \prob{\bs{\tau} \mid  \bs{\tau}_{\rm obs}} =& \int_{\bs{n_e}} \prob{ \bs{n_e}, \bs{\tau} \mid  \bs{\tau}_{\rm obs}} \, \mathrm{d}\bs{n_e}\\
    =& \int_{\bs{n_e}} \prob{ \bs{n_e} \mid  \bs{\tau}_{\rm obs}} \prob{ \bs{\tau} \mid  \bs{n_e}, \bs{\tau}_{\rm obs}} \, \mathrm{d}\bs{n_e},
    \label{eq:jointcond_b}
\end{align}
where in the second line we used the product rule of probability distributions \citep{kolmogorov1956}.
By working through Eqs.~\ref{eq:jointcond_a} and \ref{eq:jointcond_b}, we discover that if $\prob{\bs{\tau} \mid \bs{n_e}} = \prob{ \bs{\tau} \mid  \bs{n_e}, \bs{\tau}_{\rm obs}}$ is true, then our method is equivalent to first inferring FED and then using that distribution to calculate $\Delta \mathrm{TEC}$.
In Appendix~\ref{app:dsep} we prove that the expressions in Eqs.~\ref{eq:jointcond_a} and \ref{eq:jointcond_b} are equal due to the linear relation between FED and $\Delta\mathrm{TEC}$  because the sum of two Gaussian RVs is again Gaussian.
Most importantly, this result would not be true if $\Delta\mathrm{TEC}$ was a non-linear projection of FED.

We refer to this as implicit tomography as opposed to explicit tomography, wherein the FED distribution would be computed first and the $\Delta \mathrm{TEC}$ computed second \citep[e.g.][]{JIDLING2018141}.
This explains why our kernel is able to accurately predict $\Delta\mathrm{TEC}$ in regions without nearby calibrators. 
The computational savings of our approach is many-fold compared with performing explicit tomography, since the amount of memory that would be required to evaluate the predictive distribution of FED everywhere would be prohibitive.
Finally, the use of GPs to model ray integrals of a GP scalar field is used in the seismic physics community for performing tomography of the interior of the  Earth.

\subsection{Temporal differential TEC correlations}
One clearly missing aspect is the temporal evolution of the ionosphere. 
In this work we have considered instantaneous realisations of the FED from a spatial GP; however, the inclusion of time in the FED GP is straightforward in principle. 
One way to include time is by appending a time dimension to the FED kernel, which would mimic internal (e.g. turbulence-driven) evolution of the FED field.
%with temporal correlations and appending a time dimension to the coordinate spaces.
Another possibility is the application of a frozen flow assumption, wherein the ionospheric time evolution is dominated by a wind of constant velocity $\bs{v}$, so that $n_e(\x,t) = n_e^0(\x - \bs{v} t)$. 
Here, $n_e^0$ represents the FED at time $t = 0$, and $n_e$ is a translation over the array as time progresses. 
In modelling a real dataset with frozen flow the velocity could be assumed to be \emph{piece-wise} constant in time. % with $\bs{v}$ the velocity of the frozen flow.
We briefly experimented with frozen flow and found hyperparameter optimisation to be sensitive to the initial starting point due to the presence of many local optima far from the ground-truth hyperparameters. 
We suggest that a different velocity parametrisation might facilitate implementation of the frozen flow approach.

\subsection{Structure function turnover and anisotropic diffractive scale}
\label{sec:var_func}

\begin{figure}[h]
    \centering
    \includegraphics[width=\columnwidth]{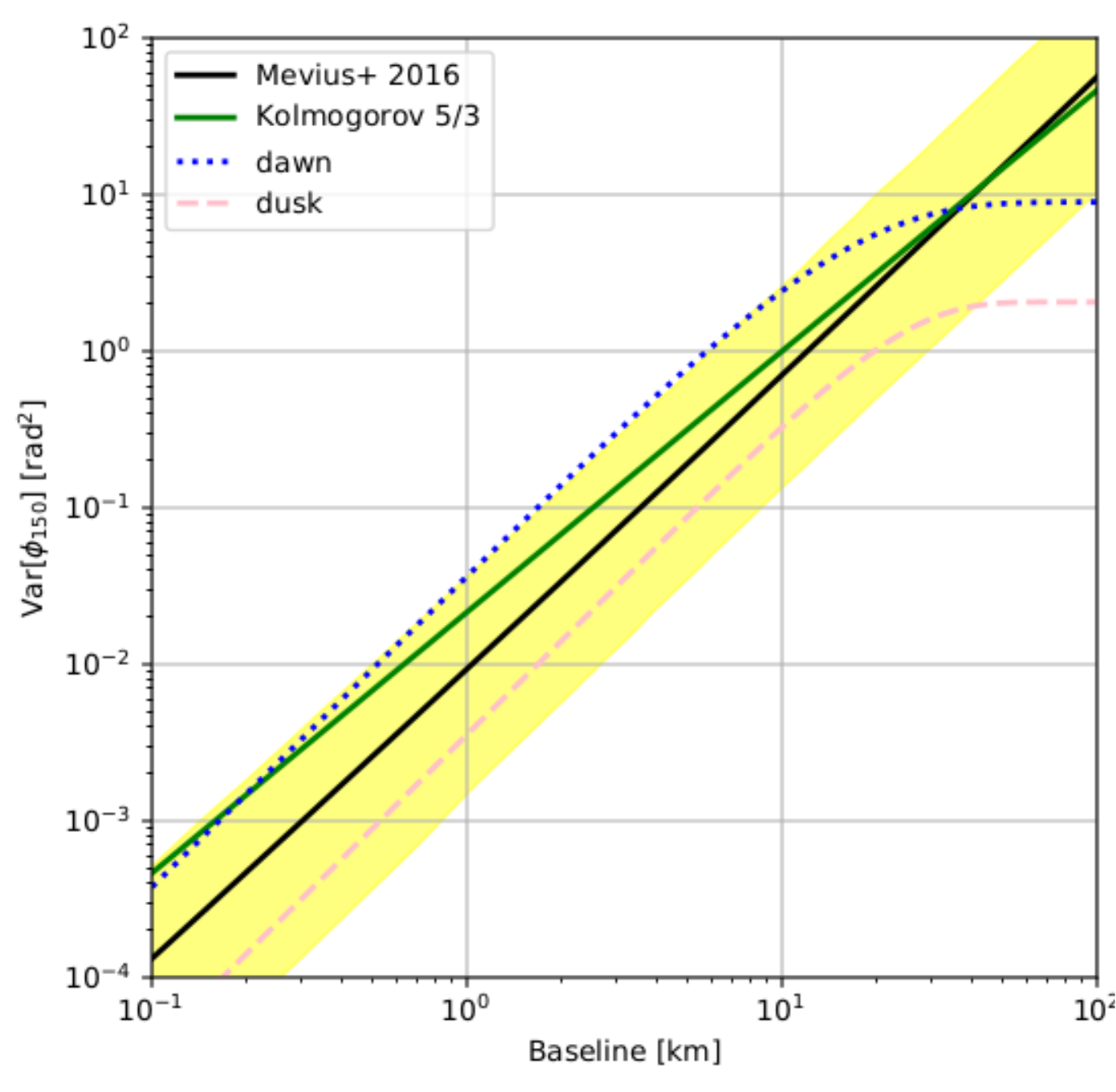}
    \caption{Structure functions predicted by our model compared with obserations and theory. The dotted and dashed lines show the phase structure function corresponding to the physical kernel, with the dawn and dusk configurations, respectively (see \textbf{Figure}~\ref{tab:varieties}). Along side is the predicted structure function of Kolmogorov turbulence with a diffraction scale of 10~km, and the structure function constrained from observations in \citet{mevius2016} with $1\sigma$ confidence region in yellow. \review{We note that \citet{mevius2016} observes a turnover, but does not characterise it, and therefore we do not attempt to plot it here.} }
    \label{fig:phase_structure}
\end{figure}

% \begin{figure}[h]
%     \centering
%     \includegraphics[width=\columnwidth]{diffractive_scale.pdf}
%     \caption{This shows the diffractive scale of the dusk ionosphere as a function of zenith angle.}
%     \label{fig:diff_scale}
% \end{figure}

The power spectrum is often used to characterise the second-order statistics of a stationary random medium, since according to Bochner's Theorem the power spectrum is uniquely related to the covariance function via a Fourier transform.
In 1941, Kolmogorov \citep[translated from Russian in][]{1991RSPSA.434....9K} famously postulated that turbulence of incompressible fluids with very large Reynolds numbers displays self-similarity.
From this assumption, he used dimensional analysis to show that the necessary power spectrum of self-similar turbulence is a power-law with an exponent of -5/3.
A convenient related measurable function for the ionosphere is the phase structure function \citep{tol2009},
\begin{align}
    D(r) =& \langle (\phi_\nu(R) - \phi_\nu(r + R))^2 \rangle_R\\
    \triangleq& \left(\frac{r}{r_{\rm diff}}\right)^\beta,
\end{align}
where the expectation is locally over locations far from the boundaries of the turbulent medium, which is often characterised by an outer scale.
The quantity $r_{\rm diff}$ is referred to as the diffractive scale, and is defined as the length where the structure function is $1~\mathrm{rad}^{2}$.
Under Kolmogorov's theory of 1941, $\beta = 5/3$.
Observations from 29 LOFAR pointings constrain $\beta$ to be $1.89 \pm 0.1$, slightly higher than predicted by Kolmogorov's theory, and the diffractive scale to range from 5 to 30~km \citep{mevius2016}. 

In \textbf{Figure}~\ref{fig:phase_structure} the structure functions of the physical kernel are shown for the dawn and dusk varieties, alongside Kolmogorov's $\beta=5/3$ and the \citet{mevius2016} observations.
Though not plotted, \citet{mevius2016} also find that there is a hint of a turnover in the structure functions they observed, which they suggest might be a result of an outer scale in the context of Kolmogorov turbulence. However, these latter authors conclude that longer baselines are needed to properly confirm the turnover and its nature.
The dawn and dusk structure functions are nearly parallel with observations, and have turnovers that result because the FED covariance functions decay to zero monotonically and rapidly beyond the HPD.
Interestingly, despite the fact that the FED kernels used for the dawn and dusk ionospheres have different spectral shapes, the structure functions have similar slopes.
The difference between the dawn and dusk structure functions can be seen in the curvature of their turnovers.

Our model provides an explanation for the observed shape of structure functions, which Kolmogorov's theory of 1941 fails to provide, namely the existence of a turnover, and a slope deviating from five-thirds.
Specifically, a turnover requires only FED correlations that are stationary, isotropic, and monotonically decreasing (SIMD).
Both the dawn and dusk ionosphere varieties experimented with predict slopes compatible with observations.
Moreover, as shown in Appendix~\ref{app:varianceFunction}, our model in conjunction with the SIMD FED kernel is falsifiable by observing a lack of plateau. 

\citet{mevius2016} also observe anisotropy in the measured $r_{\rm diff}$ as a function of pointing direction, and suggest that it is due to FED structures aligned with magnetic field lines \citep{2015GeoRL..42.3707L}.
In total, 12 out of 29 (40\%) of their observations show anisotropy unaligned with the magnetic field lines of Earth. 
We propose a complementary explanation for the anisotropy of diffractive scale, without appealing to magnetic field lines. 
Our model implies that diffractive scale monotonically decreases with zenith angle.
This is a result of the non-stationarity of the physical kernel even if the FED is stationary.

\subsection{Low-accuracy numerical integration}
\label{sec:low_acc}
The numerical integration required to compute Eq.~\ref{eq:kernel_a} is performed using the 2D Trapezoid rule.
This requires the selection of a number of partitions along the ray.
The computational complexity scales quadratically with the number of partitions chosen, and thus a trade-off between accuracy and speed must be chosen.
% We computed a reference kernel using a 4-th order Runge-Katta integrator with adaptive stepsize and were thus able to measure the accuracy of our trapizoidal approximation as a function of number of partitions.
We found the relative error (using the Frobenius norm) to be 80\% with two partitions, 20\% with three partitions, 10\% with four partitions, and 6\% with seven partitions.
After experimentation it was surprisingly found that two partitions was sufficient to beat all competitor models, and that marginal improvement occurs after five partitions.
This suggests that even a low-accuracy approximation of our model encodes enough geometric information to make it a powerful tool in describing the ionosphere.
Ultimately, we chose to use four partitions for our trials.

% \subsection{Remark on non-probabilistic methods}
% It is a common practice in astronomy to use non-probabilistic interpolation methods like splines, or parametrised basis fitting, to model data points.
% Such methods are crude, in comparison with probabilistic methods like GPs, but they are fast and easy to implement.
% One can make such methods principled and take into account uncertainty.
% In fact, such methods can be easily recast in terms of a GP regression problem, e.g. linear or polynomial regression, giving them all the benefits of a probabilistic interpretation \citep{Rasmussen:2005:GPM:1162254}.
% However, what we have shown here is that even a set of widely successful GP models impose large biased errors in their predictions due to the sparse nature of the problem.
% In the realm of precise radio interferometric calibration, it is clear that such crude methods should be cast aside henceforth, and even naive probabilistic methods given considerable thought.
% In general, we argue that if there is a physical theory for the observables of a problem, then it should be used to model the system from a probabilistic perspective.

\section{Conclusion}

In this work, we put forth a probabilistic description of antenna-based ionospheric phase distortions, which we call the physical model.
We assumed a single weakly scattering ionosphere layer with arbitrary height and thickness, and free electron density (FED) described by a Gaussian process (GP).
We argue that modelling the FED with a GP locally about the mean is a strong assumption due to the small ratio of FED variation to mean as evidenced from ionosphere models.
We show that under these assumptions the directly observable $\Delta\mathrm{TEC}$  must also be a GP. We provide a mean and covariance function that are analytically related to the FED GP mean and covariance function, the ionosphere height and thickness, and the geometry of the interferometric array.

In order to validate the efficacy of our model, we simulated two varieties of ionosphere -- a dawn (rough FED) and dusk (smooth FED) scenario -- and computed the corresponding $\Delta\mathrm{TEC}$ for the Dutch LOFAR-HBA configuration over a wide range of experimental conditions including nominal and sparse-and-noisy conditions.
We compared this physical kernel to other widely successful competitor GP models that might naively be applied to the same problem.
Our results show that we are always able to learn the FED GP hyperparameters and layer height -- including from sparse-and-noisy $\Delta\mathrm{TEC}$ data -- and that the layer thickness could likely be learned if a height prior was provided.
In general, the physical model is better able to represent observed data and generalises better to unseen data.

Visual validation of the predictive distributions of $\Delta\mathrm{TEC}$ show that the physical model can accurately infer $\Delta \mathrm{TEC}$ in regions far from the nearest calibrator.
Residuals from the physical model (0.5--1~mTECU) are smaller and less correlated than those of competitor models (2--10~mTECU).
In terms of mean equivalent source shift resulting from incorrect predictions, the physical model mean equivalent source shift is approximately half of that of  the best competitor model.
We show that our model performs implicit tomographic inference at low cost, which is because $\Delta \mathrm{TEC}$ is a linear projection of FED and the FED is a GP.
We suggest possible extensions to incorporate time, including frozen flow and appending the FED spectrum with a temporal power spectrum.
Our model provides an alternative explanation for the \citet{mevius2016} observations: phase structure function slope deviating from Kolmogorov's five-thirds, the turnover on large baselines, and diffractive scale anisotropy.

In the near future, we will apply this model to LOFAR-HBA datasets and perform precise ionospheric calibration for all bright sources in the field of view.
It is envisioned that this will lead to clearer views of the sky at the longest wavelengths, empowering a plethora of science goals.

\begin{acknowledgements}
J.\ G.\ A. and H.\ T.\ I. acknowledge funding by NWO under `Nationale Roadmap Grootschalige Onderzoeksfaciliteiten', as this research is part of the NL SKA roadmap project.
J.\ G.\ A. and H.\ J.\ A.\ R. acknowledge support from the ERC Advanced Investigator programme NewClusters 321271. R.\ J.\ vW. and M.\ S.\ S.\ L.\ O. acknowledge support of the VIDI research programme with project number 639.042.729, which is financed by the Netherlands Organisation for Scientific Research (NWO).  
M.\ S.\ S.\ L.\ O. thanks Jesse van Oostrum for helpful discussions.
\end{acknowledgements}

\bibliographystyle{aa}
\bibliography{cite}

\onecolumn
\clearpage
\appendix

% \section{The ionospheric FED as a lognormal random field: moments}
% Considering the ionospheric FED $n_e$ to be lognormally distributed at every point $\x \in \mathcal{X} \subset \mathbb{R}^3$ is a generalisation of the model adopted in the main text. From this perspective, the adopted model can be seen as the `small-fluctuation', or \emph{linear} limit of a more general setup. Although the linear limit is likely valid for the majority of observing time, it might not always. Another reason to consider this more general case is that, if $n_e$ is lognormal, the probability of realising negative values is $0$, satisfying the physical requirement that is violated when $n_e$ is assumed to be normal.\\
% However, when $n_e$ is a lognormal random field, TEC and $\Delta\mathrm{TEC}$ are not exactly GPs.
% A rough conceptual understanding that motivates this claim is the following. A $\Delta\mathrm{TEC}$ is a linear combination of TECs (namely a difference), and a TEC is a line integral over the FED. An integral is the limit of a Riemann sum, which is also a linear combination. However, linear combinations of lognormal RVs have no direct closed-form analytic expression.

\section{Derivation of tomographic equivalence}
\label{app:dsep}

We now explicitly prove the assertion that Eq.~\ref{eq:jointcond_a} is equal to Eq.~\ref{eq:jointcond_b}, that is,
\begin{align}
    \int \prob{\bs{\tau} \mid \bs{n_e}} \prob{\bs{n_e} \mid \bs{\tau}_{\rm obs}}\, \mathrm{d}\bs{n_e} = \int \prob{ \bs{n_e}, \bs{\tau} \mid  \bs{\tau}_{\rm obs}}  \, \mathrm{d}\bs{n_e}.\label{eq:assert}
\end{align}
We note that we sometimes use the notation $\mathcal{N}[a \mid m_a, C_a]$ which is equivalent to $a \sim \mathcal{N}[m_a, C_a]$.

We define the matrix representation of the DRI operator in Eq.~\ref{eq:dri_a}, $\bs{\Delta}_{*} \bs{n_e} = \{\Delta_{\x}^{\hat{\bs{k}}}n_e \mid (\x, \hat{\bs{k}}) \in \mathcal{S}_{*} \}$, and likewise let $\bs{\Delta}$ be the matrix representation over the index set $\mathcal{S}_{\rm obs}$.
Similarly, the matrix representation of the FED kernel -- the Gram matrix -- is $\bs{K} = \{ K(\x, \x') \mid \x,\x'\in \mathcal{X}\}$.
Using these matrix representation we have the following joint distribution,
\begin{align}
    \prob{\bs{n_e}, \bs{\tau}, \bs{\tau}_{\rm obs}} =& \mathcal{N}
    \left[ 
    \begin{matrix}
    \bar{n}_e\\
    0\\
    0
    \end{matrix}
    ,\ \ \ \ 
    \begin{matrix}
    \bs{K} &\bs{K}\bs{\Delta}_*^T & \bs{K}\bs{\Delta}^T\\
    \bs{\Delta}_*\bs{K} & \bs{\Delta}_* \bs{K} \bs{\Delta}^T_* & \bs{\Delta}_* \bs{K} \bs{\Delta}^T\\
    \bs{\Delta}\bs{K} & \bs{\Delta} \bs{K} \bs{\Delta}^T_* & \bs{\Delta} \bs{K} \bs{\Delta}^T + \sigma^2 \bs{I}
    \end{matrix}
    \right].\label{eq:joint_tomo}
\end{align}

Let us first work out the left-hand side (LHS) of Eq.~\ref{eq:assert}.
Because $\bs{\tau} = \bs{\Delta}_{*} \bs{n_e}$, and using standard Gaussian identities we have,
\begin{align}
    \prob{\bs{\tau} \mid \bs{n_e}} = \mathcal{N}[\underset{\bs{\Delta}_* (\bs{n_e} - \bar{n}_e)}{\underbrace{\bs{\Delta}_* \bs{K} \bs{K}^{-1}(\bs{n_e} - \bar{n}_e)}}, \underset{\bs{0}}{\underbrace{\bs{\Delta}_* \bs{K} \bs{\Delta}_* - \bs{\Delta}_* \bs{K}\bs{K}^{-1}\bs{K} \bs{\Delta}_*}}].
\end{align}
Similarly, the second distribution on the LHS is,
\begin{align}
    \prob{\bs{n_e} \mid \bs{\tau}_{\rm obs}} = 
    \mathcal{N}[
    \bar{n}_e + \bs{K} \bs{\Delta}^T (\bs{\Delta} \bs{K} \bs{\Delta}^T + \sigma^2 \bs{I})^{-1} \bs{\tau}_{\rm obs}
    , 
     \bs{K}- \bs{K} \bs{\Delta}^T (\bs{\Delta} \bs{K} \bs{\Delta}^T + \sigma^2 \bs{I})^{-1}\bs{\Delta} \bs{K}].
\end{align}
We now apply belief propagation of Gaussians \citep{6789575} to evaluate the integral on the LHS,
\begin{align}
    &\int \prob{\bs{\tau} \mid \bs{n_e}} \prob{\bs{n_e} \mid \bs{\tau}_{\rm obs}}\, \mathrm{d}\bs{n_e}\\
    =& \int 
    \mathcal{N}[\bs{\tau} \mid \bs{\Delta}_* (\bs{n_e} - \bar{n}_e), \bs{0}]
    \mathcal{N}[\bs{n_e} \mid
    \bar{n}_e + \bs{K} \bs{\Delta}^T (\bs{\Delta} \bs{K} \bs{\Delta}^T + \sigma^2 \bs{I})^{-1} \bs{\tau}_{\rm obs}
    , 
    \bs{K} - \bs{K} \bs{\Delta}^T (\bs{\Delta} \bs{K} \bs{\Delta}^T + \sigma^2 \bs{I})^{-1}\bs{\Delta} \bs{K}]
    \, \mathrm{d}\bs{n_e}\\
    =& \mathcal{N}[
    \underset{\bs{\Delta}_* \bs{K} \bs{\Delta}^T (\bs{\Delta} \bs{K} \bs{\Delta}^T + \sigma^2 \bs{I})^{-1} \bs{\tau}_{\rm obs}}
    {\underbrace{-\bs{\Delta}_* \bar{n}_e + \bs{\Delta}_* (\bar{n}_e +  \bs{K} \bs{\Delta}^T (\bs{\Delta} \bs{K} \bs{\Delta}^T + \sigma^2 \bs{I})^{-1} \bs{\tau}_{\rm obs})}}
    , 
    \bs{\Delta}_* \bs{K} \bs{\Delta}_*^T - \bs{\Delta}_* \bs{K} \bs{\Delta}^T (\bs{\Delta} \bs{K} \bs{\Delta}^T + \sigma^2 \bs{I})^{-1} \bs{\Delta} \bs{K} \bs{\Delta}^T_*
    ].
\end{align}

In order to work out the right-hand side (RHS), we simply condition Eq.~\ref{eq:joint_tomo} on $\bs{\tau}_{\rm obs}$ and then marginalise $\bs{n_e}$ by selecting the corresponding sub-block of the Gaussian,
\begin{align}
    &\prob{\bs{n_e}, \bs{\tau} \mid \bs{\tau}_{\rm obs}} ,\\
    =& \mathcal{N}
    \left[
    \left(
    \begin{matrix}
    \bar{n}_e\\
    0
    \end{matrix}\right) + \left(\begin{matrix}
    \bs{K}\bs{\Delta}^T\\
    \bs{\Delta}_*^T \bs{K} \bs{\Delta}^T
    \end{matrix}
    \right)
    \left(\bs{\Delta} \bs{K} \bs{\Delta}^T + \sigma^2 \bs{I}\right)^{-1} \bs{\tau}_{\rm obs}
    ,
    \left(
    \begin{matrix}
    \bar{K} &\bs{\Delta} \bs{K} \bs{\Delta}^T_*\\
    \bs{\Delta}_* \bs{K} \bs{\Delta}^T & \bs{\Delta}_* \bs{K} \bs{\Delta}^T_*
    \end{matrix}\right) - \left(\begin{matrix}
    \bs{K}\bs{\Delta}^T\\
    \bs{\Delta}_* \bs{K} \bs{\Delta}^T
    \end{matrix}\right)
    \left(\bs{\Delta} \bs{K} \bs{\Delta}^T + \sigma^2 \bs{I}\right)^{-1} \left(\begin{matrix}
    \bs{\Delta}\bs{K} &
    \bs{\Delta} \bs{K} \bs{\Delta}^T_*
    \end{matrix}\right)
    \right]
\end{align}
Marginalising over $\bs{n_e}$ is equivalent to neglecting the sub-block corresponding to $\bs{n_e}$.
Therefore, the RHS is,
\begin{align}
     \int \prob{ \bs{n_e}, \bs{\tau} \mid  \bs{\tau}_{\rm obs}}  \, \mathrm{d}\bs{n_e} =& \mathcal{N}
    \left[
    \bs{\Delta}_* \bs{K} \bs{\Delta}^T \left(\bs{\Delta} \bs{K} \bs{\Delta}^T + \sigma^2 \bs{I}\right)^{-1} \bs{\tau}_{\rm obs}
    ,
    \bs{\Delta}_* \bs{K} \bs{\Delta}^T_*
    - 
    \bs{\Delta}_* \bs{K} \bs{\Delta}^T \left(\bs{\Delta} \bs{K} \bs{\Delta}^T + \sigma^2 \bs{I}\right)^{-1}
    \bs{\Delta} \bs{K} \bs{\Delta}^T_*
    \right].
\end{align}
$\blacksquare$

\section{Derivation of the differential TEC variance function and its limits}
\label{app:varianceFunction}
We derive the $\Delta\mathrm{TEC}$ variance function $\sigma_{\Delta\mathrm{TEC}}^2(d)$ for zenith observations ($\bs{k} = \bs{k}' = \hat{\bs{z}}$) by considering a baseline between an antenna-of-interest at $\x_i = \x_j$ and a reference antenna at $\x_0 = \bs{0}$. To use the Pythagorean theorem later, we assume that this baseline lies in the plane of the local horizon, i.e. perpendicular to the zenith. Without loss of generality, we can orient the coordinate axes such that this baseline lies along the $\hat{\x}$ direction, so that $\x_i - \x_0 = d\hat{\x}$. Here $d \triangleq ||\x_i||$ is the distance between the two antennae. We then take the general covariance function $K_{\Delta\mathrm{TEC}}\left(\left[\vec{x}_i, \vec{x}_0, \hat{\vec{k}}\right], \left[\vec{x}_j, \vec{x}_0, \hat{\vec{k}}'\right]\right)$, and find that in this particular case
\begin{align}
    \sigma_{\Delta\mathrm{TEC}}^2(d)
    &\triangleq K_{\Delta\mathrm{TEC}}\left(\left[\vec{x}_i, \vec{x}_0, \hat{\vec{z}}\right], \left[\vec{x}_i, \vec{x}_0, \hat{\vec{z}}\right]\right)\\
    &= \sum_{p_1 = 0}^1 \sum_{p_2 = 0}^1 \left(-1\right)^{p_1 + p_2} \int_{0}^{b} \int_{0}^{b} K_{n_e}\left(||\vec{x}_{\left(1 - p_1\right)i} - \vec{x}_{\left(1 - p_2\right)i} + \hat{\vec{z}} \left(s_1 - s_2\right) ||\right)\ \mathrm{d}s_1\mathrm{d}s_2,
\end{align}
where $K_{n_e}$ is an arbitrary stationary and isotropic kernel (such as the Exponentiated Quadratic and Mat\'ern $\frac{3}{2}$ kernels considered earlier) for the FED.
The two terms where $p_1$ and $p_2$ are equal give the same contribution, as do the two terms for which $p_1$ and $p_2$ are unequal. By subsequently applying the Pythagorean theorem in this last case (i.e. $p_1 = 0$ and $p_2 = 1$, and vice versa), we find
\begin{align}
\sigma_{\Delta\mathrm{TEC}}^2(d) &= 2 \int_0^b \int_0^b  K_{n_e}\left(\left| s_1 - s_2 \right|\right) - K_{n_e}\left(\sqrt{d^2 + \left(s_1 - s_2\right)^2}\right) \mathrm{d}s_1 \mathrm{d}s_2.
\end{align}
We manipulate this result to obtain a more insightful expression. First, we note the (implicit) presence of three parameters with dimension length: ionospheric thickness $b$, reference antenna distance $d$, and FED kernel half-peak distance $h$. We perform transformations to dimensionless coordinates $u_1 = \frac{s_1}{h}$ and $u_2 = \frac{s_2}{h}$ to reveal that the \emph{shape} - though not the absolute {scale} - of the function $\sigma_{\Delta\mathrm{TEC}}^2(d)$ is governed only by the length-scale ratios $\frac{b}{h}$ and $\frac{d}{h}$, and the particular functional form of $K_{n_e}$.\\
Furthermore, for stationary covariance functions, we have $K_{n_e} = \sigma_{n_e}^2 C_{n_e}$, where $C_{n_e}$ is the corresponding dimensionless correlation function.\\
These considerations enable us to express the $\Delta\mathrm{TEC}$ structure function as a dimensionless, shape-determining double integral appended by dimensionful prefactors; i.e.
\begin{align}
    \sigma_{\Delta\mathrm{TEC}}^2(d) = 2 \sigma_{n_e}^2 h^2 \int_0^\frac{b}{h} \int_0^\frac{b}{h} C_{n_e}\left(h\left|u_1 - u_2\right|\right) - C_{n_e}\left(h\sqrt{\left(\frac{d}{h}\right)^2 + \left(u_1 - u_2\right)^2}\right) \mathrm{d}u_1 \mathrm{d}u_2.
\label{eq:varianceDTEC}
\end{align}
We first note that the variance of $\Delta\mathrm{TEC}$ is simply proportional to the variance of $n_e$.
Secondly, we note that $h\left|u_1 - u_2\right| < h\sqrt{\left(\frac{d}{h}\right)^2 + \left(u_1 - u_2\right)^2}$ for any non-zero $d$, so that $C_{n_e}\left(h\left|u_1 - u_2\right|\right) > C_{n_e}\left(h\sqrt{\left(\frac{d}{h}\right)^2 + \left(u_1 - u_2\right)^2}\right)$ for all monotonically decreasing correlation functions $C_{n_e}$ (or, equivalently, covariance functions $K_{n_e}$). With the integrand always positive, we see that the {integral} must be a strictly increasing function of $\frac{b}{h}$ (which occurs in the integration limits). Therefore, we conclude that for stationary, isotropic, and monotonically decreasing (SIMD) FED kernels with HPD $h$, the $\Delta\mathrm{TEC}$ variance increases monotonically with the thickness of the ionosphere $b$. Simply put: thicker SIMD ionospheres cause larger $\Delta\mathrm{TEC}$ variations.\\\\
Let us now consider three limits of the $\Delta\mathrm{TEC}$ zenith variance function, that all do not require $K_\mathrm{FED}$ to decrease monotonically.
In the short-baseline limit, i.e. $\frac{d}{h} \to 0$, we have $C_{n_e}\left(h\sqrt{\left(\frac{d}{h}\right)^2 + \left(u_1 - u_2\right)^2}\right) \to C_{n_e}\left(h\left|u_1 - u_2\right|\right)$. We therefore find that $\sigma_{\Delta\mathrm{TEC}}^2 \to 0$ irrespective of other parameters, recovering that the variance of $\Delta\mathrm{TEC}$ vanishes near the reference antenna.
In the long-baseline limit, i.e. $\frac{d}{h} \gg \frac{b}{h} > 1$, we see that $\sqrt{\left(\frac{d}{h}\right)^2 + \left(u_1 - u_2\right)^2} \approx \frac{d}{h}$, since $\left(u_1 - u_2\right)^2 < \left(\frac{b}{h}\right)^2 \ll \left(\frac{d}{h}\right)^2$. 
Assuming $C_{n_e}(d) \approx 0$ when $\frac{d}{h} \gg 1$, the integrand reduces to $C_{n_e}\left(h\left|u_1 - u_2\right|\right) - C_{n_e}\left(h \cdot \frac{d}{h}\right) \approx C_{n_e}\left(h\left|u_1 - u_2\right|\right)$. We find that in this case,
\begin{align}
    \sigma_{\Delta\mathrm{TEC}}^2 \approx 2\sigma_{n_e}^2 h^2\int_0^\frac{b}{h} \int_0^\frac{b}{h} C_{n_e}\left(h\left| u_1 - u_2 \right|\right) \mathrm{d}u_1 \mathrm{d}u_2.
\label{eq:plateau}
\end{align}
This is the plateau value of the $\Delta\mathrm{TEC}$ variance that our model predicts for the long-baseline limit.
    
Another way to arrive at the plateau value expression of Equation~\ref{eq:plateau} is by considering the statistical properties of $\mathrm{TEC}$ first. 
In a computation analogous to the one for $\Delta\mathrm{TEC}$ in Section~\ref{sec:model}, one can derive the general $\mathrm{TEC}$ covariance function $K_\mathrm{TEC}$. 
The variance of $\tau_i^{\hat{\bs{z}}}$ (the TEC of antenna $i$ while observing towards the zenith $\hat{\bs{z}}$) is straightforwardly shown to be
\begin{align}
    \mathbb{V}\left(\tau_i^{\hat{\bs{z}}}\right) = \sigma_{n_e}^2 h^2\int_0^\frac{b}{h} \int_0^\frac{b}{h} C_{n_e}\left(h\left| u_1 - u_2 \right|\right) \mathrm{d}u_1 \mathrm{d}u_2.
\label{eq:varianceTEC}
\end{align}
We highlight the absence of a dependence on $i$ at the RHS. As a $\Delta\mathrm{TEC}$ is simply a $\mathrm{TEC}$ differenced with a $\mathrm{TEC}$ for a reference antenna observing in the same direction, we have
\begin{equation}
    \sigma_{\Delta\mathrm{TEC}}^2 = \mathbb{V}\left(\tau_i^{\hat{\bs{z}}} - \tau_0^{\hat{\bs{z}}}\right) = \mathbb{V}\left(\tau_i^{\hat{\bs{z}}}\right) + \mathbb{V}\left(\tau_0^{\hat{\bs{z}}}\right)
,\end{equation}
where the second equality only holds when the TECs are independent. This is exactly the scenario considered in the long-baseline limit. Plugging in Equation~\ref{eq:varianceTEC} recovers the plateau level.
We can find a general upper bound to the variance of $\Delta\mathrm{TEC}$ in terms of physical parameters. To this end, we note that the integrand in Equation~\ref{eq:varianceDTEC} is maximised when, over the full range of integration, the value of the first term is 1 whilst the second term is equal to the infimum of the correlation function. Calling $\inf_\mathbb{R}\ \{C_{n_e}(r): r \in \mathbb{R}_{>0}\} \triangleq I$, we find the inequality,
    \begin{align}
        \sigma_{\Delta\mathrm{TEC}}^2 \leq 2\sigma_{n_e}^2 h^2 \int_0^\frac{b}{h}\int_0^\frac{b}{h} 1 - I\ \mathrm{d}u_1 \mathrm{d}u_2 = 2 \left(1 - I\right) \sigma_{n_e}^2 b^2.
    \end{align}
For strictly positive FED kernels that decay to zero at large distances (such as the EQ and Mat\'ern kernels considered in this work), we  find $\sigma_{\Delta\mathrm{TEC}}^2 \leq 2\sigma_{n_e}^2b^2$. Kernels resulting in anticorrelated FEDs produce the constraint $\sigma_{\Delta\mathrm{TEC}}^2 \leq 4\sigma_{n_e}^2b^2$ or tighter. By measuring $\sigma_{\Delta\mathrm{TEC}}(d)$, one can bound the product $\sigma_{n_e}b$ from below. The strongest bound is obtained for large $d$.

% \section{Distribution of source shift}

% Consider two antennas placed separated along East-West by a baseline of $b$, and consider a source at hour-angle angle $\phi$. 
% From purely geometrical considerations, by introducing an extra path-length of $\delta s$ to one antenna, the source will appear to shift by,
% \begin{align}
%     \delta l =& \arcsin{\left(\frac{\delta s}{b} + \sin{\phi}\right)} - \phi\\
% \end{align}
% Consider a stochastic process that randomly introduces path-length perturbations over the course of a time interval according to $\delta s \sim \mathcal{N}[0, \sigma_s^2]$.
% Let us calculate the mean and variance of the source shift over this time interval, under the assumption that $\frac{\delta s}{b} \ll 1$,
% \begin{align}
%     \mathbb{E}[\delta l] \approx& \mathbb{E}\left[\frac{\delta s}{b \cos{\phi}}\right]\\
%     =& 0\\
%     \mathbb{E}[\delta l^2] \approx& \mathbb{E}\left[\frac{\delta s^2  \sec^2\phi}{b^2}\right]\\
%     =& \frac{\sigma_s^2  \sec^2\phi}{b^2}
% \end{align}
% In terms of $\Delta\mathrm{TEC}$ errors, $\delta \tai$, we have that $\frac{\delta s}{b} = \frac{q^2}{\epsilon_0 m_e \nu^2 b}\delta \tau \approx 1.16\arcsec \left(\frac{b}{10\mathrm{km}}\right)^{-1}\left(\frac{\nu}{150\mathrm{MHz}}\right)^{-2}\left(\frac{\delta \tau}{\mathrm{mTECU}}\right)$
\end{document}